\documentstyle[preprint,aps]{revtex}   
\newcommand{\nc}{\newcommand}
\nc{\be}{\begin{equation}}
\nc{\ee}{\end{equation}}
\nc{\ra}{\rightarrow}
\nc{\gam}{\gamma \gamma}
\nc{\br}{\eta \ra \pi^0 \gam}
\nc{\bb}{\bibitem}
\nc{\omg}{\omega}
\nc{\g}{\gamma}
\nc{\mor}{\omega \rho}
\nc{\pa}{\partial}
\nc{\rw}{$\rho\!-\!\omega\,$}

\def\hhha{\rule[-3.mm]{0.mm}{7.mm}}
\def\hhhb{\rule[-3.mm]{0.mm}{9.mm}}
\def\hhhc{\rule[-3.mm]{0.mm}{3.mm}}
\def\hhhd{\rule[-3.mm]{0.mm}{2.mm}}
\def\hhhu{\rule[-3.mm]{0.mm}{12.mm}}
\def\hhhv{\rule[-3.mm]{0.mm}{9.mm}}
\begin{document}
\tightenlines    
\draft           
\preprint{\vbox{\null \hfill    Eur.Phys.J. C22 (2001) 503-520\\
                                        \null \hfill SLAC--PUB--8827 \\
                                        \null \hfill LPNHE 2001--04  \\
                                        \null\hfill nucl--th/0107047 }}
\title{Isospin Symmetry Breaking within the HLS Model:\\
A Full ($\rho$, $\omg$, $\phi$) Mixing Scheme}
\author{M.~Benayoun$^{a,b}$ and 
H.B. O'Connell$^c$ \footnote{Supported by the US Department of Energy
under contract DE--AC03--76SF00515}
        } 
\address{$^a$ CERN, Laboratoire Europ\'een pour
la Recherche Nucl\'eaire, 1211 Gen\`eve 23, Switzerland}
\address{$^b$ LPNHE des Universit\'es Paris VI 
et VII--IN2P3, Paris, France}
\address{$^c$ Stanford Linear Accelerator Center, Stanford University,
         Stanford CA 94309, USA}
\date{Submitted July 18th, 2001, Revised October 3rd, 2001}
\maketitle
\begin{abstract}
We study the way isospin symmetry violation
can be generated within the Hidden Local Symmetry (HLS)  Model. We show that
isospin symmetry breaking effects on pseudoscalar mesons naturally
induces correspondingly effects within the physics of vector mesons, 
through kaon loops.  In this way, one recovers all 
features traditionally expected from $\rho-\omg$ mixing and one 
finds support for the Orsay phase modelling of the $e^+e^- \ra
\pi^+ \pi^-$ amplitude. We then examine an effective procedure
which generates mixing in the whole $\rho$, $\omg$, $\phi$ sector
of the HLS Model. The corresponding model allows 
us to account for
all two body decays of light mesons accessible
to the HLS model in modulus and phase, leaving aside
the $\rho \ra \pi \pi$ and $K^* \ra K \pi$ modes only,
which raise a specific problem. Comparison with
experimental data is performed and covers modulus and
phase information~; this represents 26 physics quantities
successfully described with very good  fit quality
within a constrained model which accounts for SU(3) breaking,
nonet symmetry breaking in the pseudoscalar sector and, now,
isospin symmetry breaking.

\end{abstract}

\newpage

\section{Introduction}

\indent \indent
Despite the large difference in the $u$ and $d$ current quark masses,
isospin violation in the strong interaction is typically at the order
of a few percent, such as the $\pi^\pm - \pi^0$ mass difference.
This is because the scale is set not by $(m_u-m_d)/(m_u+m_d)$
but $(m_u-m_d)/m_s$ \cite{Gross:1979ur}. Interest in the 
contribution of isospin violation is therefore usually confined to systems
where both theoretical (or at least phenomenological)
and experimental precision are high; for example
$a_\mu$ \cite{Cirigliano:2001er}, CP violation in $B\to2P$ 
(where $P\equiv$ pseudoscalars) and other CKM-matrix systems
\cite{Grossman:1999av,Gardner:1999wb,Gardner:1999gz,Fuchs:2000hq},
the pion form-factor \cite{Kubis:2000db,O'Connell:1997mx}
and various aspects of charge symmetry violation in the $NN$
system \cite{Machleidt:2001vh}.

However, in $e^+e^- \to \pi^+\pi^-$ the isospin violating process of
\rw mixing produces a large effect on the interaction. This is due to both
the isospin independence of the initial vertex (the coupling of the
$\omega$ to the photon is only a third of the coupling of the
$\rho^0$ to the photon) and 
narrow width of the $\omega$ (in the region of the $\omega$ resonance the
cross-section is approximately 40\% larger than it would be without
\rw mixing). 
Therefore any strongly interacting system where the $\rho^0$ and 
$\omega$ have significant (if not necessarily large) production amplitudes
can expect a similar enchancement in $\pi^+\pi^-$ pair production
in the \rw interference region.
Lipkin realised this would apply to various decays in the $B$ system
\cite{Lipkin:1995hn}. Building on this,
Enomoto and Tanabashi discovered a decay channel
that would show a sizeable direct CP asymmetry,
$B^-\to\rho^-(\rho^0/\omega) \to \rho^- \pi^+\pi^-$. Here the 
penguin term exists only for $B^-\ra \rho^- \omega$ not
the $B^-\ra \rho^- \rho^0$ and the necessary
penguin/tree interference arises through \rw mixing
with the strong phase courtesy of the $\omega$ propagator
\cite{Enomoto:1996cv} (for further details see Ref.~\cite{Gardner:1998yx}).
This gives a renewed interest to the description of isospin symmetry breaking.

Having just said that \rw mixing can lead to large effects, it is important
to explain the quoted figure ($\sim2\%$) for the $\omega \to 2\pi$
branching fraction. The pion form factor can be defined (and a definition
is a useful thing) through \cite{Gardner1}
\be
F_\pi(s)=F_\rho(s)\left(1+\frac{f_{\omega\gamma}}{f_{\rho\gamma}}
\frac{\tilde{\Pi}_{\rho\omega}}{s-m_\omega^2+im_\omega\Gamma_\omega}
\right).
\ee
Though the mixing amplitude $\tilde{\Pi}_{\rho\omega}\simeq -4300$ MeV$^2$ 
is small
compared with the scale of $m_\omega^2$ the extremely narrow 
$\Gamma_\omega=8.4$ MeV allows the isospin violating contribution to
be sizeable. Correspondingly the $\omega\to 2\pi$ decay must pass through the
$\rho^0$ and thus the attenuation factor is 
$\tilde{\Pi}_{\rho\omega}/m_\rho\Gamma_\rho$ and so is down by an order
$\Gamma_\omega/\Gamma_\rho\simeq 0.05$. 

This question of scales and the meaning of the vector  meson resonance
states themselves must
be firmly kept in mind when considering the effects of \rw mixing (or
indeed any isospin violation). We shall see that this has a recent 
application.

\vspace{0.5 cm}

With this respect, it is useful to introduce isospin symmetry breaking
within a context especially designed in order to account for physics of
light vector and pseudoscalar mesons simultaneously and fully. The framework
of vector meson dominance (VMD) models is certainly the most appropriate
and, among them, the Hidden Local Symmetry (HLS) Lagrangian, with its non--anomalous 
\cite{HLS} and anomalous sectors \cite{FKTUY} is a good candidate, taking
into account its phenomenological success.

In this paper we extend our previous work on symmetry breaking within this context
by readdressing  first the $\rho-\omega$ mixing \cite{Renard:1972aj,O'Connell:1994uc}
and then the  full $\rho-\omega-\phi$ mixing. From our previous studies,
we already know how  SU(3) symmetry breaking has to be introduced  \cite{heath}
close to lines first proposed in Ref.~\cite{BKY}~; in order to yield
an appropriate description of physics information for decay processes
involving $\eta$ and $\eta'$ mesons, it has been shown that 
nonet symmetry in the pseudoscalar (PS) sector should also be broken~;
a way was proposed in \cite{rad} which provides a good understanding
of all radiative decays of light mesons. Slightly later \cite{chpt}, 
we showed that this way of breaking nonet symmetry in the PS
sector can be derived from Chiral Perturbation Theory.

However, tree level amplitudes are not sufficient in order to
account for the physics of vector mesons. As clear from the observed shape
of the pion form factor \cite{barkov,cmd2}, pion loop effects 
can hardly be neglected when describing the $\rho$ meson
\cite{ff,ff1,ff2,Gardner2,klingl}. Without a $\omg-\phi$
mixing mechanism, any description of their decay modes
becomes definitely poor. This is traditionally introduced by means 
of a mixing angle. It has been shown that kaon loop effects
are the simplest mechanism within HLS  for generating
$\omg-\phi$ mixing \cite{mixing}. These loop effects 
can be accomodated within the HLS model in an effective way,
by introducing  vector meson self--masses and (loop) transition
amplitudes  like $\omg \leftrightarrow \phi$ within the Lagrangian.
In this way one generates the appropriate pion form factor shape
and all corrections to $\omg/\phi$ decays \cite{mixing}.

However, effects of isospin symmetry breaking (like 
$\omg/\phi \ra \pi^+ \pi^-$, for instance) still remain outside
the HLS framework. We shall show that the loop mechanism
which generates the $\omg-\phi$ mixing gives a handle to
introduce isospin symmetry breaking also by considering
loop effects, mainly kaon loops. This mechanism, together
with the U(3)/SU(3) symmetry breaking procedure  recalled above,
will be shown to provide a clear understanding of (almost) 
all decay modes accessible to a VMD approach.

The outline of the paper is as follows.
In Sections \ref{sect2}, \ref{sect3} and   \ref{sect4} we 
define a mechanism for isospin symmetry breaking  and
examine its consequences on the $\rho-\omg$ sector in isolation. 
In Section \ref{sect5}, we illustrate with the pion form factor the 
consistency of this approach~; we thus show that the HLS model, 
broken as we propose, gives to the pion form factor its
well known expression in terms of the so--called Orsay phase.  

Having shown that this approach is consistent, we extend it
in Section \ref{sect6} to a scheme involving the full $\rho-\omega-\phi$ 
sector of the HLS Model.  We describe in this Section
how the coupling constants for all two body decays ($VPP$, $VP \gamma$,
$VVP$, $P\gamma\gamma$, $V e^+e^-$) can be derived from elementary 
information provided. In this approach, the  $\rho-\omg-\phi$ mixing
appears to be the at the origin of the  $\omg/\phi \ra \pi \pi$ decay
processes, which are described both in modulus and phase.

In Section \ref{sect7} we apply this model to fit all data
related to VMD, except for $\rho \ra \pi \pi$ and $K^* \ra K \pi$ 
which settle a specific problem, not examined here. The picture
obtained is impressively successful. Finally, we conclude in Section
\ref{conclusion}. We give in the Appendix most formulae and in three
Tables most of our results which cover 26 physics quantities
simultaneously fitted within a unified framework.

\section{Physical Fields or Ideal Fields}
\label{sect2}
\indent \indent
In this Section and in the following one, we concentrate on the $\rho-\omg$ 
mixing in isolation. This allows to outline the method we 
use in order to construct a full mixing scheme for vector mesons.

Within a context of an effective field model where we definitely stand,
one can  assume without any loss of generality that the
$\rho-\omg$ mixing is produced by an effective Lagrangian term
of the form
\be
{\cal L}_{mixing}= \Pi_{\mor}(s) \rho_I \omg_I
\label{one}
\ee

The use  of such a term for parametrising
$\rho-\omg$ mixing is  a usual assumption (for a review see 
\cite{ff,ff1}, where a thorough discussion of the origin, properties 
and values of $\Pi_{\mor}(s)$ can also be found). 
Its  origin within the HLS model is related
to kaon loop effects which introduce a correction of order
${\cal O}((m_{K^+}^2-m_{K^0}^2)/m_{K}^2) \sim (m_u-m_d)/m_s$, as will
be discussed in Section \ref{sect4} below. 

We have denoted above $\rho_I$ and $\omg_I$  the ideal field combinations 
({\it i.e.}, non--strange pure  isospin states)~; the corresponding physical 
fields will be denoted $\rho$ and $\omg$. For the present purpose, 
and following general ideas \cite{ff,ff1}, we
only need to assume that $\Pi_{\mor}(s)$ is a real analytic
function of $s$ ({\it i.e.} $\Pi_{\mor}^*(s)=\Pi_{\mor}(s^*)$
where the symbol $~^*$ denotes the usual complex conjugation).
 
Whatever its origin, the term in 
Eq. (\ref{one}) plays by modifying the $\rho-\omg$ mass term by adding a 
non--diagonal piece, as for the kaon loop effects responsible
for the $\omg-\phi$ mixing \cite{mixing}. Among the possible origins
of the term in Eq. (\ref{one}), pion loop effects  have been considered. 
Although this specific contribution is ruled out within the HLS model 
\cite{tony} -- which rather previleges kaon loop effects -- it
has been studied in bi-local field models \cite{Mitchell:1994jj,Mitchell:1997dn}.
Following the Renard argument \cite{Renard:1972aj}, 
the expected sizeable contribution to the imaginary part of $\Pi_{\rho\omega}$ 
from the pion loop is cancelled by a direct $\omega \ra 2\pi$ term.

\subsection{Loop Effects and Mixing}

\indent \indent
There is a consistent way to account for leading loop effects within the HLS
context~;  this turns out to modify the mass term in the Lagrangian
by including all vector meson self--energies and transition amplitudes
like $\phi_I \ra \omg_I$, as discussed in \cite{mixing}, but also{\footnote{
At the same order, one might have to include also  the parent $\rho_I \ra \phi_I$
transition amplitude, as will be seen in Section \ref{sect4}. 
}} $\rho_I \ra \omg_I$. In this approach, loop effects 
are considered by their effects at tree level only
through modified vector meson masses.

In this case, the relevant piece of the effective Lagrangian, 
quadratic in the fields, is given by~:
\be
\displaystyle
{\cal L}=\frac{1}{2} 
\{~[m^2+\Pi_\rho(s)]~ \rho_I^2 + [m^2+\Pi_\omg(s)] ~\omg_I^2
+2\Pi_{\mor}(s)~\rho_I \omg_I ~\}
\label{two}
\ee

Following the approach developed for the $\omg-\phi$ mixing \cite{mixing}, 
we have introduced the $\rho$ and $\omg$  self--energies as given
in Ref. \cite{mixing}~; these are real analytic function of $s$. 
We have also assumed that $\rho_I$ and $\omg_I$ have the same (Higgs--Kibble, HK) mass
$m$  ~; one could depart from 
this by starting with a SU(2)$\times$ U(1) symmetry \cite{UGM} instead of SU(3)
and thus break the $\rho -\omg$ mass generacy already at tree level. 
However, we have preferred  here 
neglecting  isospin breaking effects on HK masses, as it is a side issue for the
problem under study. Within the HLS model, the common $\rho -\omg$ mass is 
\cite{HLS} $m^2=af_\pi^2g^2$ in terms of the HLS parameter $a$,
of the universal vector coupling $g$ and the pion decay constant
$f_\pi$.

The diagonalization procedure of the HLS Lagrangian 
with one loop corrections is presented in detail in
Ref. \cite{mixing}.
We simply recall that the desired diagonalization
is obtained by performing the following linear field tranformation~:
\begin{equation}
\left(
   \begin{array}{ll}
     \rho \\[0.5cm]
     \omg \\
   \end{array}
\right)
=
\left(
   \begin{array}{llll}
        ~~\cos{\delta}(s) & \sin{\delta} (s)  \\[0.5cm]
	-\sin{\delta}(s) & \cos{\delta}(s)  \\
    \end{array}
\right)
\left(
   \begin{array}{ll}
     \rho^I \\[0.5cm]
     \omg^I   \\
   \end{array}
\right)
\label{three}
\end{equation}
which connects the physical fields to their ideal combinations.
It leads to physical fields which behave like analytic
functions of $s$~; this can be interpreted as a non--local
effect which could be expected as we are not dealing with 
fundamental (quark and gluon) degrees of freedom. 
Additionally, it implies at tree level (in this approach)
analytical shapes which fit well with  physics observations~;
for instance, the broad shape of the $\rho$ meson propagator
which shows up in the pion form factor is generated here mostly 
by the pion loop agregated to the $\rho_I$ mass term (see Eq. (\ref{two})).
 
The angle $\delta(s)$ -- possibly complex --
should be chosen in such a way that the mixed $\rho \omg$ term, 
appearing still in Eq. (\ref{two}) after the change to physical
fields, identically vanishes. This provides{\footnote{
It should be noted that the denominator in Eq. (\ref{four})
is nothing but the difference of the $\rho_I$
and $\omg_I$ {\it effective} running masses as they occur in
the  Lagrangian Eq. (\ref{two}).}}~:
\be
\tan{2\delta}(s) = \displaystyle\frac{2 \Pi_{\mor}(s)}
{\Pi_{\rho}(s)-\Pi_{\omg}(s)}
\label{four}
\ee
and does not depend on the 
difference of $\rho_I$ and $\omg_I$ HK masses, as this vanishes
identically in the approximation where we stand. Moreover, the $s$ dependence 
of the $\rho-\omg$ mixing exhibited by Eq.(\ref{four}) is a property already 
considered  \cite{ff,ff1,Goldman:1992fi}.  This $s$ dependence should be expected,
as the mixing function $\Pi_{\mor}(s)$ should vanish at $s=0$ 
\cite{O'Connell:1994uc} in any
model where the vector mesons couple to conserved currents.

\subsection{Analytic Properties of the Angle $\delta(s)$}

\indent \indent
As noted in \cite{mixing} for the purpose of $\omg-\phi$ mixing in 
isolation,  angles like $\delta(s)$ above are  not  real 
for any real $s$. In fact, as is clear from Eq. (\ref{four}),
$\sin{\delta(s)}$, $\cos{\delta(s)}$ are real analytic functions{\footnote{
Actually, they might only be meromorphic in the physical sheet,
as  $\Pi_{\rho}(s)-\Pi_{\omg}(s)$ might have zeros
in the physical sheet, at $s=0$ for instance.}} 
of $s$ in an analyticity domain with the same branch point singularities
as the various self--energies or transition amplitudes~; additional
algebraic branch points may occur at odd order zeros or poles of the expression
in Eq. (\ref{four}). For our purpose, one only needs to make
weak assumptions which ensure that Eq. (\ref{three}) can be
inverted as an analytic matrix function~: we assume that
the analyticity domain contains the upper and lower lips of the 
physical region $ \{ s > 4 m_\pi^2 \}$ and that both lips can be
connected with each other by a continuous path while staying
inside this domain. This implies that a segment of $\{ s < 4 m_\pi^2 \}$ 
should also belong to this analyticity domain. 

The function $\delta(s)$,  itself, can have logarithmic singularities~; 
however, it never appears as such in the expressions we have to handle.

\subsection{Interaction Terms and Self-Energies}

\indent \indent
For completeness, and in order to  fix notations for coupling constants, 
let us recall  the relevant interaction  piece of the HLS Lagrangian
as given in \cite{mixing} in terms of renormalized fields~: 
\begin{equation}
\begin{array}{lll}
{\cal T}_1= & \displaystyle -\frac{iag}{4} Z ~[\omega_I 
+\sqrt{2} \ell_V \phi_I]~
\left [ K^+ \stackrel{\leftrightarrow}{\pa} K^- 
+ K^0\stackrel{\leftrightarrow}{\pa}\overline{K}^0 \right ] \\[0.3cm]
 ~&-\displaystyle\frac{iag}{4}\rho_I
\left [ Z \left (  K^+ \stackrel{\leftrightarrow}{\pa} K^- - 
K^0\stackrel{\leftrightarrow}{\pa}\overline{K}^0 \right)
+2 \pi^+\stackrel{\leftrightarrow}{\pa}\pi^- \right ]
\end{array}
\label{AAA}
\end{equation}

We do not introduce here loop corrections to vertices, as
there is no compelling evidence in favor of observable
effects of these with the present data accuracy, even for 
the $e^+e^- \ra \pi^+ \pi^-$ cross section \cite{ff2}  known
nowadays with very good accuracy over a wide range of invariant mass
\cite{barkov,cmd2}.
 
This Lagrangian piece depends on two SU(3) breaking parameters,
generated by the BKY breaking mechanism \cite{BKY,heath}
$Z =[f_\pi/f_K]^2 \simeq 2/3$ and $\ell_V$ fit as $\simeq 1.4$
(see Refs. \cite{rad,mixing}).

 Up to anomalous contributions we neglect (they were estimated
 negligible in \cite{mixing}),
 the $\rho_I$, $\omg_I$ and $\phi_I$ self-energies are~:
\begin{equation}
\begin{array}{ll}
\displaystyle \Pi_{\rho}(s)= 
2 g_{\rho_I K \overline{K}}^2 \Pi(s)+g_{\rho_I \pi \pi}^2 \Pi'(s)\\[0.5cm]
\displaystyle \Pi_{\omg}(s) = 
2 g_{\omega_I K \overline{K}}^2 \Pi(s)\\[0.5cm]
\displaystyle \Pi_{\phi}(s) = 2 g_{\phi_I K \overline{K}}^2 \Pi(s)\\[0.5cm]
\end{array}
\label{BBB}
\end{equation}
(see Eqs. (D1)  and (D4) in \cite{mixing}) in terms of $\Pi(s)$ and $\Pi'(s)$, 
respectively the generic kaon and pion loops \cite{mixing}, {\it i.e.} the loops 
amputated from their coupling constants to vector mesons. The coupling 
constants can be read off Eq. (\ref{AAA}) and obviously fulfill 
$|g_{\rho_I K \overline{K}}|=|g_{\omega_I K \overline{K}}|$.

Let us also recall that the Lagrangian derived after the change of fields
fulfills the condition of hermitian analyticity, as in the case of 
the $\omg-\phi$ mixing in isolation \cite{mixing}.
 
\section{The $\rho-\omg$ Mixing ``Angle"}
\label{sect3}

\indent \indent
Using Eqs. (\ref{BBB}), Eq. (\ref{four}) becomes
\be
\tan{2\delta}(s) = \displaystyle
\frac{2 \Pi_{\mor}(s)}
{g_{\rho_I \pi \pi}^2 \Pi'(s)}
\label{five}
\ee

It is quite an interesting feature of the HLS model that the difference 
between the $\omg_I$ and $\rho_I$ self-energies is the pion loop 
to which only $\rho_I$ couples.

In order to estimate this denominator in the $\rho-\omg$ peak mass region, 
one should also keep in mind that this is practically the difference of the
(complex and $s$--dependent) $\rho$ and $\omg$ square masses
as they occur in the one--loop corrected Lagrangian. 
As the width of the $\omg$ meson is negligible compared to those of the
$\rho$ meson, this gives an input for the real part of the
pion loop in Eq. (\ref{five}) valid in the neighborhood of  the $\rho-\omg$ peak.
Indeed, in this region, the real part can be identified with the difference 
of the (Breit--Wigner)  masses squared as given in the RPP \cite{PDG}. 
Indeed, these are defined  as the energy point where the real part of the 
corresponding propagators goes to zero, or equivalently where the phase
goes through $\pi/2$. Therefore, writing~:
\be
 g^2_{\rho_I \pi \pi} \Pi'(s)=R(s)-iI(s),
\label{CCC}
\ee
(see Eqs. (A8) in \cite{mixing}), 
we have {\it locally}, using the RPP masses{\footnote{
Actually, $R(s)$ is a function of $s$ which contains logarithms
and a subtraction polynomial \cite{klingl}, minimally of the form 
$\lambda s$,
with $\lambda$ real to be fixed by  means of appropriate
renormalization conditions. The Breit--Wigner formulation
turns out to approximate locally this real part by a constant
which corresponds to the mass given in the RPP \cite{PDG}.
We call this mass definition {\it observed} mass and
keep in mind that it may have little to do with the masses
as they occur in Lagrangians.}}~:
\be
R(s \simeq m_\rho^2) \simeq m_\rho^2 -  m_\omg^2 \simeq  
-(1.1 \div 1.9) ~10^{-2} {\rm GeV}^2
\label{DDD}
\ee
depending on the definition used for the {\it observed} $\rho$ mass\cite{PDG}~;
this has to be compared with the imaginary part 
\be
I(s \simeq m_\rho^2) \simeq  m_\rho \Gamma_\rho = 0.12 ~{\rm GeV}^2
\label{DDD1}
\ee

Therefore, in the mass region of the $\rho$ and $\omg$ mesons,
the real part of $\Pi'(s)$ is negligible and then
the denominator in Eq. (\ref{four}) locally reduces to its 
imaginary part with a good approximation~; additionally, $I(s)$ 
is positive there, as can be inferred from its explicit expression 
\cite{mixing}.

\vspace{1.cm}

Within the HLS model, $\Pi_{\mor}(s)$ arises naturally as the {\it difference} of 
both (neutral and charged) kaon loops  \cite{tony}. These are perfectly defined 
analytic functions \cite{mixing} of $s$ and each contains a subtraction polynomial
which should be different for neutral and charged kaon loops, at least
in order to account for isospin symmetry breaking for pseudoscalar mesons.
This issue will be discussed in some more detail in the next Section, but for the present
purpose it is enough to remark that, even neglecting isospin breaking
effects on masses (then, some logarithm functions cancel out identically), the HLS 
expression for $\Pi_{\mor}(s)$ is essentially {\it a real valued subtraction polynomial}
 which has to 
be determined through renormalization conditions  \cite{mixing}. Standard renormalization 
conditions \cite{mixing} imply that this polynomial is minimally of the form $c \cdot s$, 
with  real $c$, in agreement with general considerations \cite{ff,ff1,O'Connell:1994uc}. 

Thus, within the HLS model, the numerator is essentially real,
and the denominator is largely dominated by the imaginary part of the 
pion loop in the mass region of interest. Using the coupling constants 
which can be read off from Eq. (\ref{AAA}) above,
and writing  $c s$ the amplitude for $\Pi_{\mor}(s)$ amputated from the 
coupling constants to $\omg_I$ and $\rho_I$, we have~:
\be
\tan{2\delta}(s) \simeq \displaystyle i
\frac{Z^2}{2} \frac{c s}{I(s)}
\label{six}
\ee
Therefore, in contrast with the customary mixing case of $\omg-\phi$,
the mixing angle is close to being purely imaginary in the mass region of
interest (the $\rho-\omg$ peak value). 

\vspace{0.5cm}

The mixing scheme presented here is not in contradiction
with more standard formulations in terms of a perturbation
parameter (see \cite{ff} for instance, or more recently \cite{ROPmixing}). 
However, writing it as a complex  angle makes the connection with the $\omg-\phi$
mixing more transparent. Indeed, the nature -- real or complex --  of these mixing 
angles follows from peculiarities which could look like kinematical 
accidents, essentially the relative  values of meson masses which
determine the $s$--regions where the imaginary part of loops are non--zero.

Using the RPP world average mass and 
width values for the $\rho$ and $\omg$ mesons, a better local
approximation than Eq. (\ref{six}) for the mixing ``angle" can be written 
\be
\displaystyle \tan{2\delta}(s\simeq m_\rho^2) 
= d ~m_\rho^2~  \exp{\left \{ i \left[\pi -\arctan{
\frac{m_\rho \Gamma_\rho-m_\omg \Gamma_\omg}{m_\omg^2-m_\rho^2}} \right]\right \} }
\label{sixb}
 \ee
where $d$ is a real constant to be fit. The explicit phase
(referred to below as $\varphi$) becomes
{\footnote{If we use for $\rho$ parameters the values
given in the entry ``$\tau~e^+e^-$"  of the RPP\cite{PDG}, instead of the world
average which is somewhat less secure, this angle
value becomes $95.5 \pm 0.8$ degrees. Therefore a non--negligible
systematic error ($5^\circ$, about $7 \sigma_{stat}$) can affect this number.}}~:

\be
\displaystyle 
 \varphi={\rm Arg}[ \tan{2\delta}(s\simeq m_\rho^2)]=  100.7 \pm 0.7 ~{\rm degrees}
\label{sixc}
\ee
and does not account for a possible negative sign in the fit value for $d$.
We will see shortly that this value has to be compared with the so--called
Orsay phase frequently fit within the pion form factor in the timelike region~;
among recent fit values, let us quote the value obtained within the HLS 
framework
 \cite{ff2} $104.7^\circ \pm 4.1^\circ$.

\section{SU(2) Breaking within the HLS Model}
\label{sect4}

\indent \indent
A priori, a straightforward way to introduce isospin symmetry breaking
within the HLS model could be through the BKY mechanism \cite{heath,BKY} 
proved successful  when analyzing SU(3) symmetry 
breaking effects for radiative decays of light mesons 
\cite{Durso:1987eg,rad} or the 
properties of the $\eta-\eta'$ system \cite{chpt}. 
Actually, such an attempt has been already considered in the context of 
radiative decays of light mesons \cite{Hashimoto}.

\vspace{1.cm}

Another solution is naturally proposed by the HLS model in close
correspondence with the $\omg-\phi$ mixing. Let us name provisionally
$\ell(K^+K^-)$ and $\ell(K^0 \overline{K}^0)$ the kaon loops
amputated from coupling constants to external vector meson lines.
These loops are given by dispersion relations \cite{mixing}
which should be subtracted minimally twice in order that
the dispersion integrals converge. 

This gives rise in both cases
to a first degree polynomial in $s$ with real coefficients in order
to satisfy usual analyticity properties 
(this is discussed in some detail in Appendix A
of Ref.~\cite{mixing}); let us denote by
$P_\pm(s)$ and $P_0(s)$ resp., the subtraction polynomials associated
with the kaon loops just referred to above. Their coefficients are a priori
arbitrary and should be fixed by means of appropriate
renormalization conditions. The constant term is always chosen 
to vanish in theories where  vector mesons couple to conserved 
currents \cite{O'Connell:1994uc} as in the HLS model~; the same effect ensures 
the masslessness  of the photon. If isospin is conserved, the first
degree terms of both polynomials should clearly be equal.
However, if SU(2) is broken, there is no longer any
reason for this requirement to be made.

Therefore, breaking of SU(2) symmetry can be implemented
by having different renormalization conditions for $P_\pm(s)=c_\pm~s$ and $P_0(s)=c_0~s$.
Allowing $c_\pm \ne c_0$ appears to be a consistent effective way to break isospin
symmetry within the HLS model at one--loop order.

For clarity, let us denote by $\ell(K^+K^-)+P_\pm(s)$ and 
$\ell(K^0 \overline{K}^0)+P_0(s)$, the full kaon loops,
exhibiting this way the (free) subtraction pieces.
Up to inessential coefficients related with vector coupling 
constants and SU(3) breaking effects, we have \cite{tony}~:
\be
\left \{
\begin{array}{ll}
\displaystyle \Pi_{\phi_I \omg_I}(s) \simeq & 
\displaystyle \ell(K^+K^-)+\ell(K^0 \overline{K}^0) + P_\pm(s)+P_0(s)\\[0.5cm]
\displaystyle \Pi_{\rho_I \omg_I}(s) \simeq & 
\displaystyle \ell(K^+K^-)-\ell(K^0 \overline{K}^0) + P_\pm(s)-P_0(s)\\[0.5cm]
\displaystyle \Pi_{\rho_I \phi_I}(s) \simeq & 
\displaystyle \ell(K^+K^-)-\ell(K^0 \overline{K}^0) + P_\pm(s)-P_0(s)
\end{array}
\right .
\label{D2}
\ee

Then, quite generally, the HLS model at one loop allows for transition
among the ideal combinations of {\it all three} neutral vector mesons. It should be
remarked that these transitions are associated with kaon loops rather
than with the pion loop{\footnote{In order to be complete, we
recall that anomalous terms produce loop effects like $P\gamma$ 
or $VP$ loops which contribute to the transition amplitudes~; these have been
estimated to be numerically small\cite{mixing}.}}. It is also interesting to note that, 
even if one neglects the $K^\pm - K^0$ mass difference, the
transition amplitudes  $\Pi_{\rho_I \omg_I}(s)$ and $\Pi_{\rho_I \phi_I}(s)$
do not drop out, even if their imaginary parts identically vanish. 
Moreover, as loops are analytic functions
of $s$, real  for real $s$ smaller than the loop threshold, the 
transition amplitude  $\Pi_{\rho_I \omg_I}(s)$ is certainly real
in the region of the $\rho-\omg$ peak  (up to anomalous loop effects). 
Additionally, it certainly fulfills $\Pi_{\rho_I \omg_I}(s=0)=0$.
The order of magnitude of $\Pi_{\rho_I \omg_I}(s)$ and $\Pi_{\rho_I \phi_I}(s)$
can be derived from their imaginary parts. By expanding these expressions
in the neigborhood of the $\omg/\phi$ masses, the dominant term can be written as
$[m_{K^+}^2-m_{K^0}^2]/m_{K}^2 ~(=[m_u-m_d]/m_s)$ 
corrected by a $3/2~m_{K}^2/m^2_{\omg/\phi}$ factor.

Therefore, the HLS model allows to have naturally
a quasi real $\Pi_{\rho_I \omg_I}(s)$ in the $\omg-\rho$ peak region,
as obtained from fits \cite{ff,Gardner1,Gardner2}.
This illustrates that loop effects can be used as the main mechanism
in order to break isospin symmetry by allowing different renormalization
conditions to different kaon loops. Stated otherwise, isospin symmetry 
breaking in the pseudoscalar sector already induces corresponding effects
in the vector sector.

Moreover, one observes that the HLS model at one loop, predicts that the 
full mixing pattern concerns all three neutral vector mesons and establishes 
the $\rho_I-\phi_I$ mixing as the physics mechanism for the $\phi \ra 2 \pi$ 
decay. 

\section{The Pion Form Factor}
\label{sect5}

\indent \indent
In order to compute the pion form factor in the timelike region,
the relevant piece of the interaction Lagrangian, before
changing to physical vector fields{\footnote{We still skip
in this Section mixing with the $\phi$ meson.}}, is~:

\begin{equation}
\begin{array}{ll} 
\cal{L}= \cdots & -\displaystyle i\left [\frac{ag}{2}\rho_I + e (1-\frac{a}{2})A 
\right] \cdot
[\pi^+\stackrel{\leftrightarrow}{\pa}\pi^-]
- a e f_{\pi}^2 g \left[\rho_I +\frac{1}{3}\omega_I
\right] \cdot A 
+\cdots
\end{array}
\label{Lagem}
\end{equation}
where $A$ is the electromagnetic field, $e$ the unit electric charge,
$g$  the universal vector coupling constant and $a$ the intrinsic HLS
parameter fit to $2.35 \div 2.45$ \cite{ff2,rad,mixing}. This Lagrangian piece is
not affected by SU(3)/U(3) symmetry breakdown.

After the change to physical fields given by Eq. (\ref{three}), it
is obvious that SU(2) symmetry breaking generates a direct coupling of $\omg$ 
to $\pi^+ \pi^-$. Denoting by $g_{\rho \pi \pi}^0=\frac{ag}{2}$ the 
unbroken coupling of $\rho_I$ to a pion pair, the coupling
constants for physical $\rho$ and $\omg$ are~:
\be
\begin{array}{llll}
\displaystyle 
g_{\rho \pi \pi}= g_{\rho \pi \pi}^0 \cos{\delta(s)}~~,&
g_{\omg \pi \pi}= - g_{\rho \pi \pi}^0\sin{\delta(s)}
\end{array}
\label{CCC2}
\ee

As  $ \delta(s)$ is close to purely imaginary,
this leaves the broken $\rho$ coupling close to real and
the generated coupling of $\omg$ close to purely imaginary{\footnote{
We recall that $\sin{i\alpha}=i\sinh{\alpha}$ and $\cos{i\alpha}=\cosh{\alpha}$.}}.

For sake of conciseness, let us also define~: 
\be
\begin{array}{llll}

\displaystyle f_{\rho \g}^0=a f_\pi^2 g ~~, &
\displaystyle f_{\omg \g}^0=\frac{af_\pi^2g}{3}~~~, 
\end{array}
\label{seven}
\ee
the $\rho_I$ and $\omg_I$ couplings to a photon, 
as they come out of the standard HLS Lagrangian.

\vspace{1.cm}
Using the Lagrangian piece in Eq. (\ref{Lagem}) reexpressed in terms of physical vector fields,
it is an easy matter to compute the pion form factor. Keeping the leading terms
in $\delta(s)$, this can be written~:

\be
 F_\pi(s)= \displaystyle  1 -\frac{a}{2} 
-\frac{f_{\rho \g}^0 g_{\rho \pi \pi}^0}{D_\rho(s)}~ \cos^2{\delta}
+\frac{f_{\omg \g}^0 g_{\rho \pi \pi}^0}{D_\omg(s)}~ \sin{\delta}\cos{\delta}
\label{ten}
\ee
where the $D_{\rho/\omg}(s)$ are the inverse vector meson propagators   written
$D_V(s)=s-m_V^2+im_V \Gamma_V(s)$ in most phenomenological studies, by releasing 
the analyticity assumption~; in the one--loop Lagrangian we use, these would
essentially be written $D_V(s)=s-m^2-\Pi_V(s)$, as already obtained 
and successfully tested by \cite{klingl} on $e^+e^- \ra \pi^+ \pi^-$ data.

In order to make the correspondence with Eqs. (\ref{sixb}) and (\ref{sixc}),
and with usual formulae for the pion form factor \cite{ff2}, let us state $d/2=-A$ 
and use $\varphi$,
the phase in Eq. (\ref{sixc}). Assuming $d$ is small enough, we
also have $\tan{2\delta} \simeq \sin{2\delta}$ and we can approximate
the above expression in the neighborhood of the $\rho-\omg$ peak by~:

\be
 F_\pi(s)= \displaystyle  1 -\frac{a}{2} 
-\frac{f_{\rho \g}^0 g_{\rho \pi \pi}^0}{D_\rho(s)}
-A e^{i\varphi}\frac{f_{\omg \g}^0 g_{\rho \pi \pi}^0}{D_\omg(s)}
\label{eleven}
\ee

This is nothing but the HLS expression of the pion form factor  \cite{ff2}
expressed in terms of the so--called Orsay phase, named here $\varphi$.

So, isospin breaking expressed in terms of loop effects 
gives a consistent picture for the pion form factor
and reaches the correct Orsay phase value (see Eq. (\ref{sixb})).
Therefore, an ``imaginary angle'' occuring when breaking isospin symmetry 
is what permits to recover a quite standard and traditional formulation for the 
pion form factor. 

It is an interesting feature that the $\omg-\rho$ mixing, which expresses
isospin symmetry violation, appears in correspondence with the $\omg-\phi$
mixing, produced by the same sort of loop effects. The specific character
of the $\rho-\omg$ mixing is the dominance of the subtraction term, which carries
most of the SU(2) symmetry breaking information in our approach.

In order to be complete, one can estimate the modulus of $\delta$. 
In the vicinity of the $\omega$ meson mass, one has:
\be
\displaystyle  \left | \tan{2\delta}\right |^2 \simeq
\frac{\Gamma(\omg \ra \pi^+ \pi^-)}{\Gamma(\rho \ra \pi^+ \pi^-)} = (1.24\pm 0.17) ~10^{-3}
\label{twelve}
\ee
which corresponds to a negligible ``angle" of about 1 degree times $i$.
This is indeed very small
but quite comparable in  magnitude to the (real)   $\omg-\phi$
mixing angle (about 3 degrees).

\section{The Full Mixing Pattern}
\label{sect6}

\indent \indent
It follows from the Sections above that, basically, the mixing
pattern exhibited by the HLS model at one loop involves the full
triplet ($\rho_I$, $\omg_I$, $\phi_I$) as soon as isospin symmetry
is broken, as it is in real life. In most physics studies of
light meson decays it is usual to neglect isospin breaking effects{\footnote{See, 
however, Ref. \cite{Hashimoto} for an attempt to describe radiative decays of 
light mesons.}}. A noticeable exception 
is  the pion form factor, because of the important $\rho-\omg$ 
interference structure and of the $\rho-\phi$ interference which shows 
up through the decay mode $\phi \ra \pi^+ \pi^-$ \cite{PDG,phi1,phi2}.
An interesting account of the $\rho-\phi$ 
mixing can also be found in the recent \cite{ROPmixing} in 
connection with  the $\phi \omg \pi$ coupling.

However, from the final remarks in the Section above,
one could ask oneself whether accounting only partly
for vector meson mixing effects is legitimate. Indeed,
we have just seen that the $\omg_I-\phi_I$ mixing (measured by its 
--real-- angle)  and the $\rho_I-\omg_I$ mixing
(measured by its --imaginary-- angle) are quite comparable
in magnitude. 

\subsection{Diagonalization Procedure}

\indent \indent
When the $\phi$ field is ``switched on", the effective Lagrangian piece
quadratic in the fields changes from Eq. (\ref{two}) to~:
\begin{equation}
\begin{array}{ll}
\displaystyle
{\cal L}= &
\frac{1}{2} 
\{~[m^2+\Pi_\rho(s)]~ \rho_I^2 + [m^2+\Pi_\omg(s)] ~\omg_I^2
+[\ell_V m^2 +\Pi_\phi(s)]~ \phi_I^2 \\[0.5cm]
~ & +2\Pi_{\omg_I \rho_I}(s)~\rho_I \omg_I +2\Pi_{\omg_I \phi_I}(s) ~\omg_I\phi_I
+2\Pi_{\rho_I \phi_I}(s) \rho_I \phi_I \}~~.
\end{array}
\label{Diag1}
\end{equation}

Self-energies and transition amplitudes have been defined
in Eqs. (\ref{BBB}) and (\ref{D2}) respectively. 

In order to compute amplitudes involving  the {\it physical}
$\rho$, $\omg$ and $\phi$ mesons, Eq. (\ref{Diag1}) should be diagonalized.
This gives the physical fields as algebraic expressions in terms
of the ideal field combinations $\rho_I$, $\omg_I$ and $\phi_I$.
In these expressions the coefficients of the (linear) relations
are analytic functions of $s$, which basically depend on three
``angles" through relations much more complicated than Eq. 
(\ref{three}). 

One can obviously define three such ``angles" corresponding each to the 
case where one among $\rho_I$, $\omg_I$ and $\phi_I$  is ``switched off".
As already stated above, these ``angles" are actually (analytic)
functions of $s$ and can be real, imaginary or complex depending
on the specific $s$ value along the physical region. 

The $\omg -\phi$ mixing has been studied in isolation in \cite{mixing} and 
the corresponding mixing angle has been found real as long as $s$ is smaller 
than the two--kaon threshold~; practically, this remains true up the $\phi$ 
meson mass region.
The $\omg-\rho$ mixing angle has been considered in the previous
Sections and has been found close to purely imaginary. The third mixing 
angle describes mostly the $\rho-\phi$ mixing and  is named $\gamma$ below~;
one can easily show that its imaginary part is certainy large in the mass
region of vector meson resonances, but  a precise
estimate of its real part necessitates  assumptions on the subtraction
polynomials far beyond the scope of the present paper. 

\subsection{Transformation from Ideal to Physical Fields}

\indent \indent Therefore the general transformation we are 
interested in  is certainly linear and depends on three angles, 
each a function of $s$. 

Moreover, relying on the angle functions obtained by switching
off one among $\rho^0$, $\omg$, $\phi$, it is likely that
these angles vary little along the mass range we are interested 
in{\footnote{See Ref. \cite{mixing} for the $\omg-\phi$
mixing case in isolation.}}.
This leads us to approximate these three 
analytic functions by three constants, over the mass range
covered by the light vector and pseudoscalar mesons. 
This is a somewhat violent assumption and the ability of the model
supplied with this constraint to describe experimental data
will teach us about its validity.

This being stated, the transformation which allows to
define the physical $\rho$, $\omg$ and $\phi$ fields
in terms of $\rho_I$, $\omg_I$ and $\phi_I$ is formally
a rotation and the angles are defined by requiring
the vanishing of all mixed terms, $\rho \omg$, $\rho \phi$,
$\omg \phi$ in Eq. (\ref{Diag1}) after the change of fields.
The transformation is a real rotation -- with real angles -- 
when analytically continued below the two--pion threshold.
 
This rotation matrix can be chosen as the following 
CKM--like matrix \cite{PDG}  which was also used in order to study a 
possible glue component coupled to the $\eta-\eta'$ system \cite{rad,chpt}~:
\begin{equation}
M
=
\left[
     \begin{array}{lll}
\displaystyle \cos{\delta}\cos{\beta} & -\displaystyle \sin{\delta}\cos{\beta} & \sin{\beta}\\[0.5cm]
\displaystyle \sin{\delta}\cos{\gamma}+\cos{\delta}\sin{\beta}\sin{\gamma} &
\displaystyle \cos{\delta}\cos{\gamma}-\sin{\delta}\sin{\beta}\sin{\gamma} &
-\displaystyle \cos{\beta}\sin{\gamma}\\[0.5cm]
\displaystyle \sin{\delta}\sin{\gamma}-\cos{\delta}\sin{\beta}\cos{\gamma} &
\displaystyle \cos{\delta}\sin{\gamma}+\sin{\delta}\sin{\beta}\cos{\gamma} &
\displaystyle \cos{\beta}\cos{\gamma}\\
     \end{array}
\right]
\label{Diag4}
\end{equation}
and we define the requested field transformation by~:
\begin{equation}
\left[
     \begin{array}{ll}
     \displaystyle \omg\\[0.5cm]
     \displaystyle \rho\\[0.5cm]
     \displaystyle \phi\\
     \end{array}
\right]
=
M
\left[
\begin{array}{ll}
   \omg_I  \\[0.5cm]
   \rho_I   \\[0.5cm]
    \phi_I  
     \\
     \end{array}
\right]
~~,~~
\left[
     \begin{array}{ll}
     \displaystyle \omg_I\\[0.5cm]
     \displaystyle \rho_I\\[0.5cm]
     \displaystyle \phi_I\\
     \end{array}
\right]
=
\widetilde{M}
\left[
\begin{array}{ll}
   \omg  \\[0.5cm]
   \rho   \\[0.5cm]
    \phi  
     \\
     \end{array}
\right]
\label{Diag5}
\end{equation}

One can indeed check that $M^{-1}=\widetilde{M}$
whether $\beta$, $\delta$ and $\gamma$ are real or complex.
As stated above, the sine and cosine functions here are defined
through their underlying exponential expressions and
coincide with the standard ones for real values of their arguments. 
We recall that trigonometric functions satisfy all their known
properties, even for complex values of their arguments.

\vspace{0.5cm}

\indent \indent
If the ``angle" functions may become complex, one may wonder that
we may be violating hermiticity, as one could have rather expected
$M^{-1}= M^\dagger=\widetilde{M}^*$. This is not true, as can be seen by going 
a step prior to the approximation by constants. In this case{\footnote{
We denote by $^*$ the simple complex conjugation of matrices (with no  
transposition) and variables.}}, the relation fulfilled by $M(s)$
along the physical region can be written, using obvious notation~:
\be
M(s+i\varepsilon) M^\dagger((s+i\varepsilon)^*)=M(s+i\varepsilon) M^\dagger(s-i\varepsilon)=1
\label{angle1}
\ee
assuming that the two lips of the physical region can be connected
by a path fully contained in the physical sheet and which does not cross any cut~; 
additionally, $M(s)$ is real  below the 2--pion threshold. 
 
The sine and cosine functions defining $M(s)$ are certainly algebraic
functions of the transition amplitudes and self--energies~; therefore,
they have essentially the same branch point singularities (plus possible
additional ones we will not discuss). Therefore, $M(s)$ is certainly
a real analytic function of $s$ in a domain sketched several times
above. Then, we should have along the physical region~:
\be
M^*(s-i\varepsilon)=M(s+i\varepsilon)
\label{angle2}
\ee
which leads to 
\be
M(s+i\varepsilon) \widetilde{M}(s+i\varepsilon)=1
\label{angle3}
\ee
as has been inferred from Eq. (\ref{Diag5}). The precise 
analyticity domain where this is valid is not easy to study in the present
case~; this is, furthermore, of no consequence for the present
study{\footnote{In order that the framework of what follows holds, one
has only to {\it assume} that this analyticity domain contains a band
along Re$(s)> 4 m_\pi^2$ on both sides of the real axis and, 
connectedly, a part of the semi--axis Re$(s)~< 4 m_\pi^2$.
This working assumption does not look severe.}}.

\subsection{Radiative and Vector Decays of Light Mesons}

\indent \indent
The first important data set we shall analyze are the
radiative decays of light mesons. The Lagrangian
which allows us to derive their coupling constants can be written~:
\begin{equation}
{\cal L}_{WZW}= K \epsilon^{\mu \nu \rho \sigma}
{\rm Tr} \left [
\pa_\mu(eQA_\nu+gV_\nu)\pa_\rho(eQA_\sigma+gV_\sigma)P 
\right]
\label{Diag6}
\end{equation}
where $Q={\rm Diag}(2/3,-1/3,-1/3)$ is the quark charge matrix
and $A$ is the electromagetic field. $P$ is the pseudoscalar field 
matrix and can be found in \cite{rad} with the conventions used here.
The vector field matrix is repeated here~:
\be
V=\frac{1}{\sqrt{2}}
  \left( \begin{array}{ccc}
   (\rho_I+\omega_I)/\sqrt{2}  & \rho^+             &  K^{*+} \\
            \rho^-           & (-\rho_I+\omega_I)/\sqrt{2}    &  K^{*0} \\
            K^{*-}           & \overline{K}^{*0}  &  -\phi_I   \\
         \end{array}\label{vector}
  \right).
\label{Diag7}
\ee
in order to exhibit that the traditional field $\rho^0$
is actually the ideal isospin 1 field combination, while the physical
field associated with the $\rho^0$ meson is $\rho$ in Eqs. (\ref{Diag5}).
The coefficient $K$ in Eq. (\ref{Diag6}) is \cite{mixing} 
$K=-3/(4 \pi^2 f_\pi)$.

The various $VP\gamma$ coupling constants can be derived from
Eq. (\ref{Diag6}) in a straightforward way~; before rotating
to physical fields, they are given in the Appendix. 

It should be remarked that Eq. (\ref{Diag6}) is an expression
for the VMD assumption which connects the usual anomalous 
Wess--Zumino Lagrangian for $P \gamma \gamma$ to its VMD partner 
$VP\gamma$ through a common normalization factor ($K$). Therefore,
the treatment of $VP\gamma$ and $P \gamma \gamma$ couplings differ
only by specific symmetry breaking effects.

The physical $\rho P\gamma$, $\omg P\gamma$ and $\phi P\gamma$
couplings are easily derived using these ideal couplings and the second
Eq. (\ref{Diag5}), by collecting all contributions to the same
field combination coupling. Fully developped, they are
algebraically rather complicated, even if conceptually simple.
They can, however, be easily dealt with within a minimization program.

Let us illustrate one case and, for this purpose, write down symbolically 
a piece of  Eq. (\ref{Diag6})~:
\be
\cdots G_{\rho_I\gamma P} [\rho_I A P] + G_{\omg_I\gamma P} [\omg_I A P]
 + G_{\phi_I\gamma P} [\phi_I A P]
\label{Il1}
\ee
As symbolically, one can derive from Eqs. (\ref{Diag4}) and (\ref{Diag5}) 
three relations~:
\be
\begin{array}{lll}
\rho_I= v_{\rho_I}(\omg)\omg+ v_{\rho_I}(\rho)\rho +v_{\rho_I}(\phi) \phi
\end{array}
\label{Il2}
\ee
and the corresponding ones for $\omg_I$ and $\phi_I$ with, correspondingly,
$v_{\omg_I}$  and $v_{\phi_I}$. The three vectors just defined are simply
the columns in Eq. (\ref{Diag4}). Rewriting Eq. (\ref{Il1}) using Eq. (\ref{Il2})
and the two other parent ones, it remains only to collect all terms contributing
to $[\rho A P]$, $[\omg A P]$ and $[\phi A P]$ in order to get the coupling
constants associated with physical vector mesons.

\vspace{0.5cm}

This allows to include all radiative decay modes in our data
sample, {\it i.e.}  the possible 15 decay modes presently all measured.
Actually, the  $\pi^0 \gamma \gamma$ partial width is not used
but replaced by the pion decay constant world average 
value \cite{PDG} $f_\pi = 92.42$ MeV. 

Beside the radiative $VP\gamma$ coupling constants, the Lagrangian
Eq. (\ref{Diag6}) defines also the $VVP$ couplings. From a practical
point of view, the interesting piece derived from Eq. (\ref{Diag6}) can be 
written~:
\be
{\cal L}_1= \displaystyle
-\frac{g^2}{8 \pi^2 f_\pi} \left \{ [\omg_I \rho_I \pi^0]
+[\omg_I \rho^+ \pi^-]+[\omg_I \rho^-\pi^+] \right \}
\label{Diag8}
\ee
using obvious notations. From this, we can derive the $\phi \omg \pi^0$
coupling which allows to include the corresponding decay mode in our data
sample. The coupling constants for  $\phi \rho \pi$ \cite{phi3} 
is derived from fits to the $e^+e^- \ra \pi^+ \pi^- \pi^0$ cross section~;
the most recent fit value \cite{phi3} $g_{\phi\rho\pi}=0.815 \pm 0.021$ GeV$^{-1}$
is seemingly well established. Its parent $\omg \rho \pi$  is subject to more
controversy \cite{ROPmixing,Dolinsky,omgrhopi1,omgrhopi2} and the reported values range
between $11.7 \pm 0.5$ GeV$^{-1}$ \cite{ROPmixing} for the smallest to
$16.1 \pm 0.4$ GeV$^{-1}$ for the largest, with a prefered value \cite{Dolinsky} around
$14.3$ GeV$^{-1}$~; until clarification, it seems more secure to leave this information
outside fits and simply compare with our predictions.

Finally the relative phase of the coupling constants $\phi \omg \pi^0$
and $\omg \rho \pi$ comes from a fit to $e^+e^- \ra \omg \pi^0$~;
the most recent estimate \cite{omgrhopi2} is $-49^\circ \pm 7^\circ \pm 1^\circ$
and will be included into the data sample we shall fit.

\subsection{Information from $e^+ e^- $, $\pi^+ \pi^-$  and $K\overline{K}$ Decays}

\indent \indent
Concerning the decay of vector mesons to $e^+ e^-$, the relevant Lagrangian 
piece is \cite{rad}~:
\be
{\cal L}_{em}= \displaystyle - a e f_{\pi}^2 g \left[\rho_I +\frac{1}{3}\omega_I+
\ell_V \frac{\sqrt{2}}{3}\phi_I\right] \cdot A 
\label{Diag9}
\ee
which depends on the breaking parameter \cite{BKY,heath,rad} $\ell_V$.
It allows to derive the corresponding couplings for the physical fields
$\rho$, $\omg$ and $\phi$ using Eqs. (\ref{Diag5}) above.

The other HLS Lagrangian piece given in Eq. (\ref{AAA}) 
provides the coupling constants of the physical $\phi$ meson 
to both $K^+ K^-$ and $K_L K_S$ final state.
Finally, the $\pi^+ \pi^-$ term in Eq. (\ref{AAA}) gives the
 $\omg \pi^+ \pi^-$  and $\phi \pi^+ \pi^-$
couplings which allows us to include these partial widths
inside our data sample.

Therefore, in addition to the 14 modes, the $\phi \ra \omg \pi^0$
decay width and its phase relative to $\omg \rho \pi$,
and the $\phi \rho \pi$ coupling as stated  in the section above, we can 
add 7 more decay modes to our working data sample
($\rho /\omg /\phi \ra e^+ e^- $, $ \phi \ra K^+ K^-/K_L K_S$,
$\omg /\phi \ra \pi^+ \pi^-$).

\vspace{0.5cm}

As clear from the above Sections dealing with the $\rho -\omg$ mixing,
the phase of the $\omg$ term (denoted above $\varphi$)
relative to the $\rho$ term carries as much physics information as the 
$\omg \ra \pi^+ \pi^-$ partial width (one gives the phase of the breaking 
term, the other its modulus). Referring to the Review of Particle
Properties \cite{PDG}, there is no reported average value
and the latest fit which could have produced such information
did not include this measurement \cite{cmd2}, therefore, we
shall use the latest published fit value 
\cite{ff2} $104.7^\circ \pm 4.1^\circ$ as reference data.

There are however, former fit values for the Orsay phase
which give information on its model dependence 
\cite{Box1,Box2,Bernicha}. The reference value we choose
is somewhat median and has the virtue to reproduce the threshold
behavior predicted by Chiral Perturbation Theory with a good
accuracy \cite{ff2}~; however, from the other references just quoted,
one might conclude that systematic errors of about $10^\circ$
are not unlikely.

\vspace{0.5cm}
 
On the other hand, the fit of the $\phi \ra \pi^+ \pi^-$ rate
has been done several times \cite{PDG,phi1,phi2}, but only
one value for the phase is currently reported in the literature
\cite{phi1} and provides valuable physics information.
The reported value is $\psi=-34^\circ \pm 4^\circ$; this corresponds
to a definition where the $\phi$ inverse propagator is written
$m_\phi^2-s-im_\phi \Gamma_\phi$, opposite  in sign to the definition
currently used (see  Section \ref{sect5}). In order to recover
consistency with the rest of the information we use,
a minus sign should be absorbed in this phase which thus becomes
$\psi=146^\circ \pm 4^\circ$. 
 
These phases are the phases of the following quantities~:
\be
F_V= f_{V\gamma} G_{V \pi^+ \pi^-}~~~~, ~~V=~\rho,~\omg, ~\phi
\label{Diag10}
\ee
which are allowed to be complex for all vector mesons. 

Therefore, taking into account these 2 additional phases, 
our data sample contains 26 physics quantities; indeed,
all modes reported above, except for $\pi^0 \ra \gamma \gamma$
replaced by the world average value for $f_\pi$, the 
controversial coupling
$g_{\omg \rho \pi}$, to which we shall nevertheless compare,
as well as the phase of $G_{\rho \pi \pi}$ (unavoidably
generated by isospin breaking) which should be 
(and is found) very small.

\section{Fitting the Data Sample}
\label{sect7}

\indent \indent
We have fit the set of data listed above within the model presented 
in the above Section concerning the full mixing pattern and in the
Appendix concerning the rest of the parameter set. 

Concerning the data, all have been taken from the last issue
of the Review of Particle Properties (RPP) \cite{PDG}{\footnote{
An update of the Particle Data Table can be found at http://pdg.lbl.gov~;
some minor modifications have been made to the decay rates
considered in the present paper. They do not affect our analysis.
}}.
For data which have no existing entry in the RPP, we have chosen 
the latest reference. Therefore, among all physics quantities
which could be accessed by the model we present, only the
major modes $K^* \ra K \pi$ and $\rho \ra \pi \pi$ are
left aside, as already stated.

\subsection{Analysis of the Fit Conditions}
\label{strategies}

\indent \indent
In this Section, we aim at making clear, which are the parameters
we use and the level of freedom allowed by their existing measured value~;
afterwards, we describe the various fit strategies we have followed.

The unbroken HLS model \cite{HLS} basically depends on a very
few parameters which are not predicted and should be extracted from
data. Beside the unit electric charge $e$ which is certainly well determined,
these are the universal vector meson coupling $g$, the pion decay constant
$f_\pi=92.42$ MeV and $a$, a dimensionless parameter specific of the HLS model. 
In standard VMD models, one has $a=2$, however within the HLS model
this condition can be relaxed~; in this case,  fits to experimental data 
\cite{ours,rad,chpt,cmd2} indicate that $a = 2.3 \div 2.5$ should preferred.
This merely means that a small coupling $\gamma \pi^+ \pi^-$ survives
beside vector mesons exchanges.

Concerning symmetry breaking parameters, previous studies \cite{rad,chpt,mixing}
have already reduced the fit freedom by relating, and/or fixing the  breaking
parameters specific to the pseudoscalar sector~: we have already $Z=[f_\pi/f_K]^2=2/3$, 
as a consequence of $F_K(0)=1$ \cite{BKY,heath}~; it has been checked 
that the set of radiative and leptonic decays favors this value unambiguously \cite{rad}.
However, in view of the new result \cite{Fuchs:2000hq} on this subject, we
shall make a separate study of this parameter.

As all other models, the HLS model requires
mixing in the $\eta/\eta'$ sector, however, the two parameters involved
there (the mixing angle $\theta_P$ and the nonet symmetry breaking parameter
$x$) are algebraically related by the HLS phenomenology \cite{chpt} 
(see Eq. (\ref{rel}))  with  a high accuracy. 

In the vector sector of the HLS model, two breaking parameters, denoted
here $\ell_V$ and $\ell_T$, seem unavoidable and free, even if $\ell_V$
might be fixed sometime, when vector meson masses would be clearly understood{\footnote{
We mean by this, that the relation between theoretical masses as they occur
in Lagrangians and the corresponding measured quantities is unclear for
broad objects like $\rho$ or $K^*$. This problem
 is certainly related with the apparent difficulty to accomodate the
major decay modes of $K^*$ and $\rho$ and all other decay modes 
{\it simultaneously} within the HLS framework.}} \cite{heath,rad,mixing}.

\vspace{0.5cm}

This being stated, there remain  3 complex ``angles" (6 parameters)
which are the body of the present study. Within some approximations
(mostly, neglecting anomalous loops), one has already noted some clear guesses~:
one should be mostly real (it corresponds to the standard $\omg-\phi$
mixing angle \cite{mixing}), another close to purely imaginary (it corresponds to
the $\omg-\rho$ mixing angle examined in Sections \ref{sect3}--\ref{sect5}).
Of course, when going to numerical analysis, the validity of these guesses 
can be controlled. Moreover, the expectations just referred to have been
established above or elsewhere \cite{mixing} by considering mixing
patterns in isolation~; therefore, slight departures from these 
expectations are not unlikely.

\vspace{0.5cm}

{\bf i/}
As first attempt, we have left these 6 parameters free in the fit. We reached the 
good fit quality of $\chi^2/{\rm dof}=13.41/15$ (26 data, 11 parameters) which 
corresponds to a 57\% probability. The fit correlation matrix was
observed to exhibit large correlations between fit values 
for $\beta$ and $\gamma$ and between Re($\gamma$) and Im($\gamma$)~;
as this could well influence the fit procedure, we have looked for equivalent
parametrizations. The most appropriate we found was to use  Re($\beta$), 
Im($\beta$), Im($\gamma$) and a parameter $k$ defined by 
Re($\gamma$)=$k$ Im($\gamma$). In this case we improved the fit
quality to $\chi^2/{\rm dof}=12.59/15$, corresponding to a fit probability of 
63\%. The fit then returned $k=-0.23^{+0.40}_{-0.60}$, which indicates
that the angle $\gamma$ can be chosen imaginary ($k=0$)~; the global fit assuming
this constraint is given in the first data column of Table \ref{T1}. 

{\bf ii/}
We have explored several strategies in order to reduce the freedom in fits
by fixing several subsets of parameters. The most interesting results, with 
reasonable probabilities (above the percent level), are given in the second 
and third data columns in Table \ref{T1}. The former relies on the observation that
the $\omg - \phi$ mixing angle is well fitted real \cite{mixing}~; this leads to 
try requesting Im$[\beta$]$=0$ in order to lessen the fit freedom. The latter relies
on the observation that both  Re$[\gamma$] and Re$[\delta$], basically related
with the mixing of $\omg$ and $\phi$ to $\rho$, are quite generally yielded small
compared to the corresponding real parts~; this is true in the general framework
under examination and also in studies were these mixing phenomena
were considered in isolation. Even if somewhat brutal, these approximations 
lead both to quite reasonable fit quality.

{\bf iii/} We have also considered that there can be a functional
relation between some ``angles". From their expressions in terms of
pseudoscalar meson loops, one might guess that the ``angles" $\gamma$ and
$\delta$ could be functionally related. As we are dealing with slowly varying 
functions over the range of interest, we have tried requesting~: 
\be
\displaystyle \gamma=(\mu_1 +i\mu_2)~ \delta~~.
\label{cond2}
\ee

It happens that this relation is well accepted by the data.
The fit returned $\mu_1=(0.030^{+0.007}_{-0.006})$ and 
$\mu_2=(0.011^{+0.131}_{-0.114})~10^{-1}$ with $\chi^2/\rm{dof}=12.58/15$
(probability 63\%). 
Therefore, requiring the condition 
in Eq.~(\ref{cond2}) is certainly justified
and additionally, one gets phenomenological motivation to require $\mu_2=0$ 
from start. The corresponding fit results are displayed in the fourth data 
column of Table \ref{T1}. The fit quality reached can hardly be better.

Finally, an additional fit (not shown) assuming Im$[\beta]=0$  
and leaving free  $\mu_1$ and $\mu_2$ has been performed 
in order to test the stability of other fit parameters,
by requiring a condition expected if one interprets $\beta$
as strictly equivalent to the $\omg-\phi$ mixing angle in isolation.
The result practically coincides with the second
data column in Table \ref{T1}, including its fit quality,
and returns $\mu_1=(0.354 \pm 0.044)~10^{-1}$  and 
$\mu_2=(-0.270 \pm 0.040)~10^{-1}$. The various
contributions to the $\chi^2$ implies that this fit
and the second data column in Table \ref{T1} 
give the same description of the data with the same probability.

{\bf iv/} From the results given in Table \ref{T1}, it is clear that most
parameter values do not depend sensitively on the fit strategy considered.
As all fit qualities are especially favorable, no strategy can be privileged.
The single parameter which seems floating is Re[$\gamma$] which cannot be better
constrained before improving the accuracy of existing information for
$\phi \ra \pi \pi$ in modulus and phase,  and/or improving the phase of the 
$\phi \omg \pi$ coupling constant. Whether $\gamma$ could be removed as a whole 
has been considered with a negative answer. Indeed, performing a fit
with $\gamma=0$ leads to a quality  which becomes really poor 
($\chi^2/\rm{dof}=36.34/17$, probability 0.4\%).

It should be stressed that the information prominently affected by isospin 
symmetry breaking  represents 6 measurements ($\omg/\phi \ra \pi \pi$, 
$\phi \ra \omg \pi$ in modulus and phase), which requires in our 
approach 4 parameters  ($\delta$, $\mu_1$ and $\mu_2$/Im$[\beta]$). Therefore, 
even in this sector, the set of parameters is reasonably constrained
and only waits for more accurate data.

\subsection{Analysis of Fit Parameter Values}

\indent \indent
As first remark, it is clear that all fit parameters not connected
with vector meson mixing, (the five first lines in Table \ref{T1}),
are quite stable and their values compare well with previous attempts along 
the present  lines \cite{rad,mixing,chpt}. We note, however, the correlation 
between $\ell_V$ and $a$ which reaches $-90$\%~; this correlation is
purely numerical and reflects that the dependence upon $\ell_V$ within the 
set of coupling constants is actually a dependence upon the product $a\ell_V$. 

The value obtained for the pseudoscalar mixing angle has been discussed in 
\cite{chpt} and agrees quite well with recent estimates from lattice QCD 
\cite{ukqcd}. It has been shown in Ref. \cite{chpt} that  this angle is
(algebraically) related with the mixing angle  $\theta_8$ in favor within 
the ChPT community by a factor which can be predicted close to 2.

\vspace{0.5cm}

As stated above, the mixing ``angle" $\beta$ can be considered as
intimately associated with $\omg-\phi$ mixing. It should be noted 
that the value of Re($\beta$) varies little  when constraints are
put on other parameters. This real part is a $30 \sigma$ effect
and corresponds to an $\omg-\phi$ mixing angle of $-(3.2\div 3.5)^\circ \pm 0.11^\circ$, 
that is smaller than the ideal mixing angle, as found in \cite{rad,mixing}. 
These remarks allow to conclude that introducing isospin symmetry breaking,
as we propose does not affect sensitively the sector of radiative and
leptonic decays. For most parameters not intimately related
with isospin symmetry breaking, this follows expectations (see the first
five lines in Table \ref{T1})~; however, because of the "rotation" matrix 
structure, it was not obvious that Re$[\beta]$ could not shift by a 
few degrees, pushing the $\omg-\phi$ mixing angle slightly above its
ideal value. This is not observed, whatever the fit strategy.
 
All uncertainties in the fits are connected mainly with the values for
Im($\beta$), Re($\gamma$) and Re($\delta$). This reflects that, even though
valuable, most isospin breaking data are still of rather poor accuracy.

\subsection{Reconstruction of Physics Quantities}

\indent \indent
The fit parameter values allow to reconstruct 
branching fractions, coupling constants and phase factors
as predicted by our model. Dealing with errors is done by Monte Carlo
methods using the full covariance matrix of each fit in order to account
properly for correlations.  Let us denote $V_{ij}$ the covariance matrix 
element for parameters $x_i$ and $x_j$, by $\lambda_\alpha$ its eigenvalues
and by $a_i^\alpha$ the $i^{th}$ component of the $\alpha^{th}$ 
normalized eigenvector~; then any measured parameter $x_i$ can
be considered as a random variable given by~:
\be
\displaystyle
x_i=x_i^0 +\sum_{\alpha=1}^n  \varepsilon^\alpha ~ \sqrt{\lambda_\alpha} ~a_i^\alpha
\label{sample}
\ee
where $x_i^0$ is the central value returned by the fit and $\{\varepsilon^\alpha, ~\alpha=1, \cdots n\}$
is a set of independent gaussian random variables of zero mean and unit
standard deviation ($<\varepsilon^\alpha\varepsilon^\beta>=\delta_{\alpha \beta}$).

The fit quantities were the coupling constants for each process. 
These have been derived from the accepted branching fractions \cite{PDG}
--taking into account their accuracy-- and assuming  that the full widths 
and masses of mesons are random variables.

In order to reconstruct the physics (measured) quantities, in addition
to considering the fit parameters as correlated random variables, we
have assumed the mass and width of each vector meson as independent random variables 
with standard deviation given by the accepted errors \cite{PDG}~; instead,
all masses of pseudoscalar mesons were considered as fixed, except for the 
$\eta'$  meson. Finally, for the $\rho$ meson (charged and neutral) we have
considered the value given in the $\tau/e^+e^-$ entry of the RPP \cite{PDG}
for its mass and width. We thus follow the conclusion of the ALEPH Collaboration
who saw no difference for these parameters  -- within errors -- between the charged 
and neutral modes \cite{barate}.

\subsubsection{Radiative and Leptonic Decay Modes}

\indent \indent
We give in Table \ref{T2} the reconstructed branching fractions for
radiative and leptonic decays together with the recommended values \cite{PDG}.

It is interesting to compare these reconstructed values with previous fits 
done using the model we present, without introducing isospin symmetry breaking 
(see Table III and IV in \cite{rad} and Table 2 in \cite{chpt},
where nonet symmetry breaking and the pseudoscalar mixing angle
have been algebraically related). All changes are actually tiny,
confirming that breaking of isospin symmetry contributes little 
in this realm. However, two small changes can be noticed.

The first is that BR$(\rho^0 \ra \pi^0 \gamma$) becomes larger
than BR$(\rho^\pm \ra \pi^\pm \gamma$) by 4.7 \% and the predicted
branching fraction BR$(\phi \ra \pi^0 \gamma$) increases by 8\%.
These are clearly consequences of breaking isospin symmetry. 
Otherwise, whatever the additional conditions stated,
the general agreement of the reconstructed physics 
quantities with the data collected and averaged in the RPP can 
hardly be better.

Among the recent changes in the RPP, one should notice
the branching fraction for $\phi \ra \eta' \gamma$
which has now a central value in much better agreement with
our model prediction. On the other hand,
some new measurements have been recently reported which have
not influenced the RPP recommended values (neither our fits)
and might be commented. 

First, the new measurement  Br$(\rho^0 \ra e^+ e^-)=(4.67 \pm 0.15) ~10^{-5}$
reported by CMD-2 \cite{cmd2-ee} remains in good agreement with
our fit values. The second new measurement \cite{cmd2-eta}
Br$(\rho \ra \eta \gamma)=(3.28 \pm 0.36 \pm 0.24) ~10^{-4}$
has a higher central value in better agreemement with our
reconstructed value, as for 
Br$(\phi \ra \eta \gamma)=(1.287 \pm 0.013 \pm 0.063) ~10^{-2}$.
The third new measurement  \cite{cmd2-eta}
Br$(\omg \ra \eta \gamma)=(5.10 \pm 0.72 \pm 0.34) ~10^{-4}$
is in relatively poorer agreement with our predictions than
 the RPP mean value \cite{PDG}.

The SND Collaboration has also published new results on $\eta \gamma$ 
decays of vector
mesons \cite{snd-eta1,snd-eta2,snd-all2};  the branching fractions reported
are in good correspondence with our predictions. However, as 
for the CMD-2 result reported above, the new SND data for
Br$(\omg \ra \eta \gamma)=(4.62  \pm 0.71 \pm 0.18)~10^{-4}$
might indicate that our prediction for this mode is 
slightly too large.

As the predictions for Br$(\omg \ra \eta \gamma)$
are alike whatever the conditions on the model, this
(possible) $2 \sigma$ disagreement could point
towards a mass dependence of the mixing ``angles".

Before closing this Section, it is of relevance to comment
on a recent claim \cite{china} that isospin symmetry
breaking might be much larger in $\rho^0 \ra \pi^0 \gamma$
than anywhere else. From what has just been commented,
it is clear that a $\simeq 5 \%$ effect of isospin
symmetry breaking is well accepted by all data,
the former \cite{PDG,ours} and the recent SND datum as well 
(Br$(\rho^0 \ra \pi^0 \gamma)= (4.3 \pm 2.2 \pm 0.04) ~10^{-2}$)
\cite{snd-all}. We have checked
that the central value claimed by \cite{china}
(about  a factor of 2 in rates) cannot be reproduced in consistency 
with the rest of radiative decays.
 
\subsubsection{$\pi \pi$ Decay Modes}

\indent \indent
For the $\phi$ and $\omg$  decays to  $\pi \pi$, we have
used the recommended  branching fractions \cite{PDG}
and the phases fit resp.  by \cite{phi1} and \cite{ff2}.
Table \ref{T2} shows that they are well reproduced by
any of our fits. Taking into account the uncertainties
already quoted for the Orsay phase, we even cannot rule
out the solution given by the third data column.

The Collaboration CMD--2 has recently provided \cite{cmd2} 
Br$(\omg \ra \pi \pi) = (1.32 \pm 0.23) \% $ significantly
smaller than the recommended value $(2.21 \pm 0.30)\%$
we have used, and Br$(\phi \ra \pi \pi) = (1.60 \pm 0.49)~ 10^{-4}$
about $2 \sigma$ larger than the RPP value
$(0.75 \pm 0.14)~ 10^{-4}$. No phase measurement has
been correspondingly reported. 

It is worth commenting on the possible effects of these
new measurements. These have been examined within the
framework of our preferred fit strategy (the one reported
in the fourth data column of Table \ref{T1}).

We have first changed Br$(\omg \ra \pi \pi)$ to the new CMD--2
datum. The best fit obtained provides $\chi^2/\rm{dof}= 13.68/16$
(probability 62 \%). The parameter values and errors are
the same as in the fit reported in  Table \ref{T1}, except
that Im$\delta$ yields a reduced magnitude 
($-0.029 \pm 0.002$ becomes $-0.023 \pm 0.002$)~; on the other,
hand $\mu_1$ changes from $0.031 \pm 0.005$ to  $0.020 \pm 0.007$.
 Finally, the contribution of the Orsay
phase to the global $\chi^2$ is about 0.12 and does not change,
showing that the datum used remains consistent with the rest.

Having restored Br$(\omg \ra \pi \pi)$ to the RPP recommended
value, we have changed the datum for  Br$(\phi \ra \pi \pi)$
to the new result of CMD--2. The single significant change
with respect to Table \ref{T1} is the value of $\mu_1$
($0.031 \pm 0.005$ becomes $0.017 \pm  0.001$) and the fit
returned $\chi^2/\rm{dof}= 13.44/16$ (probability 64 \%).

Finally performing both changes simultaneously provides
a fit with $\chi^2/\rm{dof}= 14.83/16$ (probability 54\%)
with results merging the changes mentioned just above.

Therefore, even if some uncertainty remains for the
values of the $\gamma$ and $\delta$ angles, the model
exhibits  enough flexibility in order to accomodate
significant changes in some crucial data. Actually,
the two modes just commented determine almost solely 
the magnitude of isospin symmetry breaking.

It should also be noted that the changes just mentioned 
in the branching fractions do not give rise to inconsistencies
with the phases of the corresponding coupling constants we have 
used, which thus look more firmly established.
 
\subsubsection{$VVP$ Couplings and $3 \pi$ Decays}

\indent \indent
In all fit strategies and even by changing to new
data as reported just above, the information concerning
the $VVP$ processes is remarkably stable.

One should thus note  the nice agreement with the data reported by 
the SND Collaboration on the $\phi \ra \omg \pi^0$ process 
\cite{ROPmixing,omgrhopi2} both in branching fraction and
phase. 

The SND datum \cite{phi3} for $|g_{\phi \rho \pi}|$ is also 
reproduced with good accuracy. As the phase of this coupling constant
is unfortunately not reported we have no reference datum
to which our prediction could be compared. Such information
is in principle accessible from fit to $e^+ e^- \ra \pi^+ \pi^- \pi^0$
data \cite{phi3}, but the existence of a (complex) non resonant term{\footnote{
This term might account for the box anomaly, but also for high
mass resonances and this last effect seems hard to model in both
modulus and phase.}} in the amplitude renders this extraction hasardous.
It could also be accessed from $ e^+ e^- \ra \omg \pi^0$ data 
but nothing is reported in this respect \cite{omgrhopi1}.
Such information, if reliable, could have been valuable as it could 
dismiss at least one of the  fit strategies (see Table \ref{T3}).

Finally, the coupling constant $g_{\rho \omg \pi}$ is found consistent
with real and its value falls indeed in the expected range \cite{omgrhopi1}.
It is found slightly but significantly smaller than the value prefered by 
\cite{Dolinsky} (14.3 GeV$^{-1}$). Its value is however extremely
stable in all fits we attempted and looks accurate~; it should be
noted that this parameter is only marginally influenced by isospin
symmetry breaking and follows essentially from the set of radiative
and leptonic decays.

The decay rates for $\omg/\phi \ra \pi^+ \pi^- \pi^0$
are, of course, determined by $g_{(\phi/\omg) \rho \pi}$ coupling
constants {\it and} a model for the $\rho$ propagator and  
the $\rho \ra \pi \pi$ decay amplitude . Therefore, our model
can be considered as giving a good description of these,
up to effects related with modelling the $\rho$ meson 
propagator for phenomenological purposes.

\subsubsection{$\phi \ra K \overline{K}$ Decay Modes}

\indent \indent
In all attempts we have performed, a non--negligible contribution
to the $\chi^2$ comes from both $\phi \ra K \overline{K}$ decay modes.
Whatever the strategy, the charged mode contributes to the $\chi^2$
by 2.2 and the neutral mode by 1.9. However, when taking into account
all sources of errors, Table \ref{T1}, clearly shows that
the disagreement with reported data is not really dramatic.

On the other hand, it is admitted that model predictions
for $\phi \ra K^+ K^-$ have to be corrected
for Coulomb interactions \cite{phiA,phi3}, which was not done above.
It has been recently shown \cite{bramon} that there 
is a slight discrepancy between the branching fractions for charged and 
neutral decay modes (about $2\sigma$) and that, accounting for
Coulomb interactions among the (very) slow charged kaons,
increases this discrepancy to $3\sigma$.

Aware of this question, we have redone our fits by removing
$\phi \ra K^+ K^-$ from the fit data set~; in this case 
we reached a fit quality of $\chi^2/{\rm dof}=9.04/15$
(88\% probability). For symmetry, we have tried removing
instead $\phi \ra K^0\overline{K^0}$~;  
we reached a fit quality of $\chi^2/{\rm dof}=9.94/15$
(82\% probability). Trying to correct the model coupling
constant as indicated in \cite{phiA} only degraded 
the nominal fit quality. Therefore, we confirm in an 
independent way the problem raised by A. Bramon {\it et al.}
\cite{bramon}. 

In order to identify a (possible) faulty measurement, we
have redone our fits by removing both $\phi$ modes from our fits.
Focusing still on the model as given in the fourth data
column in Table \ref{T1}, we reach a fit quality of
$\chi^2/\rm{dof}=8.42/14$ (probability 87\%).
What is interesting here is to consider the $\chi^2$
distance of the measurements 
to what is predicted by our model by relying only on the rest
of the data (24 measurements). We got $\chi^2=3.76$ for
the charged mode (a $2 \sigma$ effect as pointed out by \cite{bramon}),
while the neutral mode yields $\chi^2=0.90$.
When correcting the prediction for the charged mode by the Coulomb factor 
its $\chi^2$ distance increased to $\chi^2=16.6$, a $4 \sigma$ deviation.

Therefore, we confirm the issue raised by Bramon {\it et al.}
\cite{bramon}, with an additional information~: if one among
$\phi \ra K^+ K^-$  and $\phi \ra K^0\overline{K^0}$ is faulty,
it should be the former, which seems overestimated{\footnote{
It is interesting to note that systematics on $\phi \ra K^+ K^-$
are harder to estimate than those on $\phi \ra K_L K_S$, because 
the modelling of nuclear interactions of low energy charged kaons
is not still fully satisfactory. Instead, the
signature of $K_S \ra \pi^+ \pi^-$ is much cleaner.}}. 
Indeed, if we correct the model coupling constants in order
to account for Coulomb correction (1.042 for the rate), the
global  fit quality sharply degrades ($\chi^2/\rm{dof}=21.40/16$,
probability 16 \%).

\subsection{The Value of $f_K/f_\pi$}

\indent \indent
In all fits referred to above we have fixed the ratio $f_K/f_\pi$
at the central value recommended par the Particle Data Group \cite{PDG}
($f_K/f_\pi=1.226 \pm 0.012$) and neglected its error. This corresponds to 
using $[f_\pi/f_K]^2=Z=2/3$.

Instead of leaving it fixed, we allowed this ratio to vary in all
conditions described in Subsection \ref{strategies} and reported
in Table \ref{T1}. The best fits thus obtained have  improved $\chi^2$ 
with respect to Table \ref{T1} by only $\simeq 0.5$ and have one less 
degree of freedom. The different values obtained for $Z$ never differ
by more than a per mil and can be summarized by~:

\be
\frac{f_K}{f_\pi}=1.229 \pm 0.008
\label{fkpi}
\ee
which compares quite well with the PDG reference value for this datum 
recalled above. This result 
is at $2.8\sigma$ of the recent value \cite{Fuchs:2000hq} extracted from $Ke_3$ decay,
neutron decay and nuclear 
Fermi transition data ($f_K/f_\pi=1.189 \pm 0.012$). We have introduced, as fixed,
the corresponding value for $Z=0.71$ in our fits. We never reached a probability
greater than 0.5\%. Looking at the various contributions to the global $\chi^2$,
we found that it is the whole $\phi$ sector which is the most affected. 
Considering the discussion in \cite{Fuchs:2000hq} about the inputs which lead to this 
new value for  $f_K/f_\pi$, one might think that the origin of this
inconsistency is in the nuclear or in the free neutron beta decay 
datum used. We conclude herefrom that the traditional PDG value for $f_K/f_\pi$
ratio is sharply favored by the whole set of radiative decays and an
improved fit value is given by Eq. (\ref{fkpi}).

\section{conclusion}
\label{conclusion}

\indent \indent
In previous work done with other coauthors, we focused on introducing
SU(3) symmetry breaking and nonet symmetry breaking within the framework
of the Hidden Local Symmetry Model \cite{rad}. We introduced also the
$\omg -\phi$ mixing, generated by kaon loops effects,  which does not correspond 
to any symmetry breakdown \cite{rad,mixing}. This framework, supplemented with these
symmetry breaking mechanisms has been shown to provide quite a successful
picture of all radiative and leptonic decays of light vector and
pseudoscalar mesons accessible from inside the VMD framework. We have also shown
that this framework was able to explain the main features of the $\eta -\eta'$ 
mixing phenomenon \cite{chpt} in perfect agreement with all expectations of 
Chiral Perturbation Theory (ChPT)~; this led us to get a relation
between the pseudoscalar (wave function) mixing angle, 
basically at work in VMD modelling ($\simeq -10^\circ$), and the ChPT mixing
angle recently renamed $\theta_8$ ($\simeq -20^\circ$).

In the present work, we have shown that isospin symmetry breaking
can be accounted for within an effective  HLS model by means of -- 
essentially -- kaon loop effects. In contrast with the case of  
$\omg_I -\phi_I$ mixing where both kaon loops (charged and neutral) 
come additively, in the case of $\rho_I-\omg_I$ and $\rho_I -\phi_I$ mixings,
it is their difference which occurs. Relying on the properties
of Dispersion Relations, this difference should be essentially a
polynomial in $s$ with real coefficients, which is additionally
constrained to vanish at $s=0$. We argued that this polynomial
should not be identically zero, at least to account for isospin symmetry 
breaking in the pseudoscalar sector. Indeed, when isospin symmetry 
is not broken, it is quite legitimate to choose the same renormalization
conditions for both the $K^+K^-$ and $K^0 \overline{K}^0$ loops~;
instead, when  isospin symmetry is broken, this requirement
has certainly to be relaxed.

Other mechanisms than kaon loops could also be imagined. If they play
by generating $\rho_I-\omg_I$ and $\rho_I -\phi_I$ transition amplitudes,  
the  angle formalism we presented here still applies without any change. 
However, we have shown on the pion form factor,  that all properties expected 
from isospin symmetry breaking are strikingly reproduced by the kaon loop mechanism
we advocate. We then  naturally recover all properties traditionally expected 
from the $\rho -\omg$ mixing amplitude~: $\Pi_{\rho_I \omg_I}(s)$ is 
practically real in the $\rho -\omg$ peak invariant mass region, it is 
$s$--dependent and vanishes at the chiral point. 

Moreover, we were able to derive the pion form factor
in the Orsay phase formulation from our (effective) broken
Lagrangian~; the Orsay phase was shown to be strictly equivalent
to a ``rotation" by a complex angle, additionally close to purely 
imaginary. 

Using this framework, it has been possible to extend our breaking
scheme in order to include isospin symmetry breaking. Actually,
taking into account the various orders of magnitude of the breaking
parameters and of the $\omg-\phi$ mixing, it is mathematically
safer to define a full mixing scheme involving the triplet
$\rho$, $\omg$, $\phi$ as a whole. This leads us to define
a priori  a $s$--dependent rotation matrix, depending on three
angles which can be real or complex.

We have thus formulated an effective Lagrangian model which is able to
account quite successfully  for practically all physics quantities
related to VMD~: radiative decays ($VP\gamma$, $P\gamma\gamma$) , 
leptonic  decays ($V e^+ e^-$), $VVP$ couplings, and all decays
related with isospin symmetry breaking ($\omg/\phi \ra \pi \pi$,
$\phi \ra \omg \pi$) in modulus and in phase. This represents
26 physics quantities all well reconstructed. 

It should be noted that all results we previously obtained
without introducing isospin symmetry breaking are confirmed,
including the $\eta - \eta'$ and $\omg-\phi$ mixing angles.

\vspace{1.0cm}
\begin{center}
{\bf Acknowledgements}
\end{center}
HOC was supported by the US Department of Energy under contract
DE--AC03--76SF00515. We acknowledge A. Bondar, S. Eidelman and
V. Ivantchenko (Budker Institute of Nuclear Science, Novosibirsk, 
Russia) for useful discussions, suggestions and for information
concerning the data collected with the OLYA, ND, CMD--2,
and SND  detectors. We acknowledge F. Renard (University 
of Montpellier, France) for remarks on the manuscript.
We are finally indebted to R. Forty (Cern, Geneva, Switzerland), 
J.--M. Fr\`ere (ULB, Brussels, Belgium) and Ph. Leruste (LPNHE--Paris 
VI/Paris VII, France) for several comments and suggestions which
allowed us to clarify several aspects and properties
of the model developped here.

\begin{table}[htb]
\begin{tabular}{|| c  | c  | c | c  | c ||}
\hhhb
\hhhc Fixing     &   $\rho-\phi$ Imaginary &  $\omg-\phi $  Real       &  $\omg-\rho $ Imaginary    & $\omg-\rho$ \\
\hhhc Angle      &     only                &      only                 &    and                     &   and    \\
\hhhc Properties &                         &                           &   $\rho-\phi$ Imaginary    & $\rho-\phi$ \\
\hhhc   ~~       &         ~               &     ~                     &          ~                 &     proportional \\
\hline
\hline
$g$\hhhb&
$5.651 \pm 0.017$ & $5.652 \pm 0.017$ & $5.641 \pm 0.017$  & $5.652 \pm 0.017 $\\
\hline
$\theta_P$[deg.]\hhhu & --$10.33 \pm 0.20$ & --$10.32 \pm 0.20$ 
&  --$10.33 \pm 0.20$  & --$10.34 \pm 0.20$\\
\hline
$a$ [HLS] \hhhu&
$2.517\pm 0.035$  & $2.523\pm 0.034$  & $2.485\pm 0.033$ &  $2.513\pm 0.035$ \\
\hline
\hline
$\ell_V$ \hhhu&
$1.343 \pm 0.021$ & $1.337 \pm 0.021$ &
 $1.366 \pm 0.021$  & $1.346 \pm 0.021$\\
\hline
$\ell_T$ \hhhu&
$1.231 \pm 0.052$ & $1.230 \pm 0.052$ & $1.232 \pm 0.052$   & $1.230 \pm 0.052$\\
\hline
\hline
Re[$\beta$]\hhhu&
$-0.058 \pm 0.003$ & $-0.061 \pm 0.002$ & $-0.054 \pm 0.003$   & $-0.056 \pm 0.003$\\
\hline
Im[$\beta$]\hhhu&
$-0.020 \pm 0.005$ & {\bf 0.~~}& $-0.028 \pm 0.003$   &  $-0.029 \pm 0.002$\\
\hline
\hline
Re[$\delta$] \hhhu & $(0.52 \pm 0.18)~10^{-2}$
&  $(0.54 \pm 0.19)~10^{-2}$   &{\bf 0.~~}  &  $(0.55 \pm 0.19)~10^{-2}$ 
 \\
\hline
Im[$\delta$] \hhhu& $(-0.29 \pm 0.02)~10^{-1}$
& $(-0.29 \pm 0.02)~10^{-1}$  &  $(-0.31 \pm 0.02)~10^{-1}$  & $(-0.29 \pm 0.02)~10^{-1}$
 \\
\hline
\hline
Re[$\gamma$] \hhhu&
{\bf 0.~~} & $(-0.57\pm0.15)~10^{-3}$ & {\bf 0.~~}  & $(.031 \pm .005)$~Re[$\delta$]
 \\
\hline
Im[$\gamma$] \hhhu& $(-0.96 \pm 0.18)~10^{-3}$
&$(-1.16\pm 0.16)~10^{-3}$ &$(-1.06\pm 0.18)~10^{-3}$   & $(.031 \pm .005)$~Im[$\delta$]
 \\
\hline
\hline
$\chi^2/\rm{dof}$ \hhhu & 12.88/16 &  17.06/16 &20.94/17  &12.59/16 \\
Probability  & 63 \% & 38 \% & 23\%  &  70\% \\
\end{tabular}
 
\caption{
\label{T1}
Fit results under various strategies.
Parameter values written boldface means
that they are not allowed to vary~; this translates mathematically
the fit condition given on the top of the Table. 
}
 
\end{table}

\begin{table}[htb]
\begin{tabular}{|| c  | c  | c | c  | c | c ||}
\hhhc
\hhhd Fixing     &   $\rho-\phi$     &  $\omg-\phi $         &  $\omg-\rho $    &    $\omg-\rho$     & PDG\\
\hhhd Angle      &     Imaginary     &       Real            &  $\rho-\phi$     &    $\rho-\phi$     & ~~ \\
\hhhc Properties &        only       &           only        &    Imaginary     &    proportional    & ~~\\
\hline
\hline
$\rho \ra \pi^0 \gamma~ [\times 10^{4}]$ \hhhv&
$5.36 \pm 0.12$ & $5.37 \pm 0.13$ &  $5.18 \pm 0.10 $  
&  $5.37 \pm 0.13$ & $6.8 \pm 1.7$ \\
\hline
$\rho \ra \pi^\pm \gamma~ [\times 10^{4}]$\hhhv
 & $5.13\pm 0.10$ & $5.13\pm 0.10$ 
&  $5.10\pm 0.10$   &  $5.13\pm 0.10$   & $4.5 \pm 0.5$ \\
\hline
$\rho \ra \eta \gamma~ [\times 10^{4}]$ \hhhv&
$3.18 \pm 0.08$  & $3.18 \pm 0.08$  & $3.15 \pm 0.08$ 
&  $3.18 \pm 0.08$  & $2.4^{+0.8}_{-0.9}$ \\
\hline
$\eta' \ra \rho \gamma~ [\times 10^{2}]$ \hhhv&
$33.91 \pm 3.16$ & $33.92 \pm 3.13$ &
 $33.52 \pm 3.04$  & $33.93 \pm 3.16$& $30.2 \pm 1.3$ \\
\hline
\hline
$ K^{*\pm} \ra K^\pm \gamma  [\times 10^{4}]$\hhhv&
$9.89 \pm 1.01$ & $9.78 \pm 1.01$ & $9.85 \pm 1.03$   
&  $9.89 \pm 1.01$  & $9.9 \pm 0.9$ \\
\hline
$ K^{*0} \ra K^0 \gamma  [\times 10^{3}]$\hhhv&
$2.31 \pm 0.33$ &  $2.31 \pm 0.32$   &  $2.30 \pm 0.32$ 
& $2.31 \pm 0.33$ & $2.3 \pm 0.2$ \\
\hline
\hline
$\omg \ra \pi^0 \gamma~ [\times 10^{2}]$ \hhhv&
$8.49 \pm 0.10$ & $8.49 \pm 0.10$ & $8.48 \pm 0.11$  
&  $8.49 \pm 0.10$ & $8.5 \pm 0.5$ \\
\hline
$\omg \ra \eta \gamma~ [\times 10^{4}]$ \hhhv&
$7.72 \pm 0.15$  & $7.74 \pm 0.16$  & $7.86 \pm 0.14$ 
&  $7.69 \pm 0.15$  & $6.5 \pm 1.0$ \\
\hline
$\eta' \ra \omg \gamma~ [\times 10^{2}]$ \hhhv&
$2.79 \pm 0.26$ & $2.77 \pm 0.26$ & $2.89 \pm 0.26$  
 & $2.79 \pm 0.26$& $3.03 \pm 0.31$ \\
\hline
\hline
$\phi \ra \pi^0 \gamma~ [\times 10^{3}]$ \hhhv&
$1.37 \pm 0.09$ & $1.36 \pm 0.09$ & $1.38 \pm 0.09$   
& $1.38 \pm 0.09$  & $1.26 \pm 0.10$ \\
\hline
$\phi \ra \eta \gamma~ [\times 10^{2}]$ \hhhv&
$1.29 \pm 0.02$  & $1.28 \pm 0.02$  & $1.29 \pm 0.02$ 
& $1.29 \pm 0.02$   & $1.297 \pm 0.033$ \\
\hline
$\phi \ra \eta' \gamma~ [\times 10^{4}]$ \hhhv&
$0.58 \pm 0.02$ & $0.59 \pm 0.02$ & $0.58 \pm 0.02$  
 & $0.58 \pm 0.02$& $0.67^{+0.35}_{-0.31}$ \\
\hline
\hline
$\eta \ra \gamma \gamma~ [\times 10^{2}]$ \hhhv&
$39.45 \pm 3.74$ & $39.43 \pm 3.74$ &
 $39.32 \pm 4.02$  & $39.45 \pm 3.74$& $39.33 \pm 0.25$ \\
\hline
$\eta' \ra \gamma \gamma~ [\times 10^{2}]$ \hhhv&
$2.13 \pm 0.20$ & $2.13 \pm 0.20$ &
 $2.18 \pm 0.19$  & $2.13 \pm 0.20$& $2.12 \pm 0.14$ \\
\hline
\hline
$\rho \ra e^+ e^- ~ [\times 10^{5}]$ \hhhv&
$4.70 \pm 0.16$ & $4.73 \pm 0.16$ &
 $4.54 \pm 0.15$  & $4.69 \pm 0.16$& $4.49 \pm 0.22$ \\
\hline
$\omg \ra e^+ e^- ~ [\times 10^{5}]$ \hhhv&
$6.96 \pm 0.21$ & $6.94 \pm 0.21$  &  $7.06 \pm 0.22$ 
 &  $6.97 \pm 0.21$  & $7.07 \pm 0.19$ \\
\hline
$\phi \ra ~ e^+ e^- [\times 10^{4}]$ \hhhv&
$2.96 \pm 0.04$ & $2.96 \pm 0.04$ &
 $2.96 \pm 0.04$  & $2.96 \pm 0.04$& $2.91 \pm 0.07$ \\
\hline
\hline
$\chi^2/\rm{dof}$ \hhhb & 12.88/16 &  17.06/16 &20.94/17  &12.59/16 & ~~\\
Probability  & 63 \% & 38 \% & 23\%  &  70\% & ~~\\
\end{tabular}
\caption{
\label{T2}
Reconstructed Branching fractions for radiative and leptonic decays
using the  various fit strategies. The last column displays the recommended
values from the Review of Particle Properties \protect \cite{PDG}. The 
last line gives a reminder of the fit quality given in Table \ref{T1}.
}
\end{table}

\begin{table}[htb]
\begin{tabular}{|| c  | c  | c | c  | c | c ||}
\hhha
\hhhc Fixing     &   $\rho-\phi$     &  $\omg-\phi $         &  $\omg-\rho $    &    $\omg-\rho$     & PDG\\
\hhhc Angle      &     Imaginary     &       Real            &  $\rho-\phi$     &    $\rho-\phi$     & / \\
\hhha Properties &        only       &           only        &    Imaginary     &    proportional    & Reference\\
\hline
\hline
$\phi \ra K^+ K^-  [\times 10^{2}]$ \hhhv&
$50.25 \pm 0.72$ & $50.26 \pm 0.71$ &  $50.24 \pm 0.72 $  
&  $50.22 \pm 0.73$ & $49.2 \pm 0.7$ \\
\hline
$\phi \ra K^0_S K^0_L  [\times 10^{2}]$ \hhhv&
$32.95 \pm 0.47$ & $32.95 \pm 0.47$ &  $ 32.94 \pm 0.48$  
&  $32.92 \pm 0.48$ & $33.8 \pm 0.6$ \\
\hline
\hline
$\omg \ra \pi^+ \pi^-  [\times 10^{2}]$ \hhhv&
$2.23 \pm 0.30$ & $2.19 \pm 0.29 $ &  $2.32 \pm 0.31 $  
&  $2.26\pm 0.30$ & $2.21 \pm 0.30$ \\
\hline
phase of \hhha& $$ & $$ &  $ $  &   & \cite{ours} \\
$g_{\omg \pi^+ \pi^- }$  [degr]\hhha&
$103.50 \pm 4.02$ & $106.3 \pm 3.81$ &  $92.34 \pm 0.69 $  
&  $103.40 \pm 3.88$ & $104.7 \pm 4.1$ \\
\hline
\hline
$\phi \ra \pi^+ \pi^-  [\times 10^{5}]$ \hhhv&
$7.93 \pm 1.40$ & $8.15 \pm 1.45$ &  $7.60 \pm 1.24 $  
&  $7.70 \pm 1.43$ & $7.5 \pm 1.4$ \\
\hline
phase of \hhha& $$ & $$ &  $ $  &   &  \cite{phi1} \\
$g_{\phi \pi^+ \pi^- }$  [degr]\hhha&
$146.30 \pm 3.95$ & $147.5 \pm 4.09$ &  $ 146.0 \pm 3.75$  
&  $145.95 \pm 3.93$ & $146.0 \pm 4.0$($^*$) \\
\hline
\hline
$\phi \ra \omg \pi^0  [\times 10^{5}]$ \hhhv&
$4.10 \pm 0.48$ & $3.62 \pm 0.41$ &  $4.24 \pm 0.49 $  
&  $4.18 \pm 0.49$ & $4.8 \pm 2.0$ \\
\hline
phase of \hhha& $$ & $$ &  $ $  &   & \cite{omgrhopi2}  \\
$g_{\phi  \omg \pi^0 }/g_{\omg  \rho \pi^0 }$  [degr]\hhha&
$-50.90 \pm 3.63$ & $-61.91 \pm 3.05$ &  $ -52.38 \pm 3.27$  
&  $-47.92 \pm 3.52 $ & $-49 \pm 7.07$ \\
\hline
\hline
coupling \hhha& $$ & $$ &  $ $  &   &   \cite{phi3}  \\
$g_{\phi \rho \pi^0 }$ GeV$^{-1}$ \hhhv&
$0.802 \pm 0.026$ & $0.799 \pm 0.026$ &  $0.803 \pm 0.026 $  
&  $0.803 \pm 0.026$ & $0.815 \pm 0.021$ \\
\hline
phase of \hhha& $$ & $$ &  $ $  &   & ~~  \\
$g_{\phi  \rho \pi^0 }/g_{\omg  \rho \pi^0 }$  [degr]\hhha&
$19.28 \pm 4.71$ & $0.22 \pm 0.11$ &  $ 27.98 \pm 3.53$  
&  $23.98 \pm 3.47$ & $- $ \\
\hline
coupling \hhha& $$ & $$ &  $ $  &   & (see text)  \\
$g_{\omg \rho \pi^0 }$ GeV$^{-1}$ \hhhv&
$13.14 \pm 0.09$ & $13.14 \pm 0.09$ &  $13.09 \pm 0.081$  
&  $13.14 \pm 0.09$ & $11.7 \div 16.1$ \\
\hline
\hline
$\chi^2/\rm{dof}$ \hhhb & 12.88/16 &  17.06/16 &20.94/17  &12.59/16 & ~~\\
Probability  & 63 \% & 38 \% & 23\%  &  70\% & ~~\\
\end{tabular}
\caption{
\label{T3}
Reconstructed Branching fractions from various fit
 strategies, cont'd. The last column displays the recommended
 values from the Review of Particle Properties 
\protect \cite{PDG}. The last line reminds
the global fit quality given in Table \ref{T1}. The
datum indicated by ($^*$) has been corrected
in order to absorb a minus sign (see text).
}
\end{table}
 

\newpage 

\section*{Appendices}

\indent \indent
In order to be self--contained, we collect in  this Appendix  formulae
for coupling constants and partial widths~; we do not insist much
on how U(3)/SU(3) breaking is performed in the present paper, as it is the matter of
already published work \cite{heath,mixing,rad,chpt} to which the interested reader
can refer.

\appendix

\section{Details of the Breaking Model}
\label{AA}

\indent \indent
Our framework is the HLS model and the SU(3) breaking procedure 
we follow has been defined first in \cite{BKY,heath}. 
Focusing on the anomalous sector \cite{FKTUY}, all details
can be found in \cite{rad,mixing,chpt}. Here, we mainly recall breaking 
parameter properties or values of concern for the present study. 

Breaking the non--anomalous sector of the HLS model \cite{HLS,BKY,heath}
introduces a breaking parameter $Z$ strongly associated with the pseudoscalar 
(PS) sector~; it is not a free parameter but fulfills $Z=[f_\pi/f_K]^2=2/3$.

Concerning the PS sector, we have a priori 2 additional parameters.
The first is named $x$ and its departure from 1 measures  breaking of nonet 
symmetry in the PS sector. Another parameter affecting the PS sector
is the PS mixing angle $\theta_P$ (which describes the $\eta/\eta'$ sector
in terms of mixtures of singlet and octet components) {\it or} $\delta_P$
(when one prefers referring to departures from ideally mixed states). Both
angles are used and are trivially related to each other \cite{rad}.

When studying the connection between VMD, the Wess--Zumino--Witten Lagrangian
and Chiral Perturbation Theory, it has been found \cite{chpt} that the PS mixing 
angle $\theta_P$ and the nonet symmetry breaking parameter $x$ fulfill~:
\be
\displaystyle \tan{\theta_P}= \sqrt{2} \frac{Z-1}{2Z+1} x
\label{rel}
\ee
with high accuracy (in \cite{chpt}, this relation is given in terms of $z=1/Z$).
Preliminary fits have shown that this relation is still satisfied in the present
framework without any degradation~; thus it is assumed. We remind that the mixing
angle $\theta_P$ relates to the (now) more usual ChPT mixing angle $\theta_8$
by \cite{chpt} $\theta_8 \simeq 2 \theta_P$.
Therefore, concerning the PS sector, our model depends only on one free
parameter which can be either of $\theta_P$ or $x$. We choose the former.

Associated with the vector sector and, more precisely with vector meson masses, 
another breaking parameter occurs named here $\ell_V$~; it relates with another
breaking parameter ($c_V$) \cite{BKY,heath} by $\ell_V=(1+c_V)^2$. It is a priori 
subject to fit, and thus free, as the connection between reported vector meson masses \cite{PDG}
and the corresponding masses occuring in the HLS Lagrangian is unclear \cite{mixing}.

Concerning vector mesons, another breaking parameter is necessary in order
to account for  the anomalous $K^*$ sector~; it is named \cite{rad,mixing} $\ell_T$. 
It was first considered as somewhat ad hoc \cite{rad}~; however, it has been 
shown \cite{mixing} that it strictly corresponds within VMD to a breaking 
parameter defined independently by G. Morpurgo \cite{Morpurgo} within the non--relativistic
quark model and found in agreement with low energy QCD. If the partial
width value for $K^{*\pm} \ra K^\pm \gamma$ is confirmed, this parameter looks
unavoidable within VMD~; its precise meaning is still to be understood \cite{mixing}.

A possible break up of nonet symmetry in the vector sector
has  been found previously undetectable (the parameter $y$ defined
and studied in \cite{mixing}). Preliminary fits in the present study 
confirmed this conclusion and, therefore, the parameter $y$ was set to 1 
definitely.

Thus, concerning SU(3) symmetry breaking, the HLS vector sector depends
already on 2 free parameters $\ell_V$ and $\ell_T$, independently of
mixing among the ideal field combinations $\omg_I$, $\phi_I$ and $\rho_I$
associated with neutral vector mesons. This last point is the actual subject 
of the present paper.
  
\section{Basic Coupling Constants and Patial widths}
\label{BB}
\indent \indent We give in this Section all coupling constants which
cannot be trivially read off the Lagrangian pieces given in the main text.

\subsection{Radiative Decays}
Starting from the Lagrangian in Eq. (\ref{Diag6}), and using the breaking 
procedure as defined in Refs \cite{mixing,rad,chpt}, one can compute
the coupling constants for all radiative and leptonic decays of relevance. 
Let us define~:

\begin{equation}
G=\displaystyle 
-\frac{3eg}{8 \pi^2 f_\pi}~~~,~~~ 
G'=\displaystyle -\frac{3eg}{8 \pi^2 f_K}~~~, ~~~Z= [f_\pi/f_K]^2~~.
\label{brk1}
\end{equation}

Some $VP\gamma$ coupling constants are not affected by
the parameters of isospin symmetry breaking. These are~:
\begin{equation}
\left \{
\begin{array}{lll}
G_{\rho^{\pm} \pi^{\pm} \gamma}=& &\displaystyle \frac{1}{3} G \\[0.3cm]
G_{K^{*0} K^0 \gamma}=&- & 
\displaystyle \frac{G'}{3} \sqrt{\ell_T} (1+\frac{1}{\ell_T})  \\[0.3cm]
G_{K^{*\pm} K^{\pm} \gamma}=& &
\displaystyle \frac{G'}{3} \sqrt{\ell_T} (2-\frac{1}{\ell_T}) ~~. \\[0.3cm]
 \end{array}
\right .
\label{brk3a}
\end{equation}

The $\rho_I P\gamma$ coupling constants are~:
\begin{equation}
\left \{
\begin{array}{lll}
G_{\rho_I \pi^0 \gamma}=& &\displaystyle \frac{1}{3} G \\[0.3cm]
G_{\rho_I \eta \gamma}=& &\displaystyle \frac{1}{3} G
\left[\sqrt{2}(1-x)\cos{\delta_P}-(2x+1)\sin{\delta_P}\right]\\[0.3cm]
G_{\rho_I \eta' \gamma}=& &\displaystyle \frac{1}{3} G
\left[\sqrt{2}(1-x)\sin{\delta_P}+(2x+1)\cos{\delta_P}\right]~~.
\end{array}
\right .
\label{brk3b}
\end{equation}

 The other single photon radiative modes provide the following coupling
 constants~:
\begin{equation}
\left \{
\begin{array}{lll}
G_{\omega_I \pi^0 \gamma}=& & \displaystyle G \\[0.3cm]

G_{\phi_I \pi^0 \gamma}=& & \displaystyle 0 \\[0.3cm]

G_{\omega_I \eta \gamma}=& &  \displaystyle \frac{1}{9} G  \left [
\sqrt{2} (1-x) \cos{\delta_P} -(1+2x) \sin{\delta_P} \right ]\\[0.3cm]

G_{\omega_I \eta' \gamma}=& &\displaystyle \frac{1}{9} G \left [
(1+2x)\cos{\delta_P} +\sqrt{2}(1-x)\sin{\delta_P} \right ]\\[0.3cm]

G_{\phi_I \eta \gamma}=& -&\displaystyle \frac{2}{9} G \left [
Z(2+x)\cos{\delta_P}  - \sqrt{2}Z(1-x)\sin{\delta_P} \right ]\\[0.3cm]

G_{\phi_I \eta' \gamma}=&- &\displaystyle \frac{2}{9}G \left [
\sqrt{2}Z(1-x)\cos{\delta_P} + Z(2+x)\sin{\delta_P} \right ]~~.
\end{array}
\right .
\label{brk4}
\end{equation}

\subsection{ $P\gamma \gamma$ and $V-\gamma$ Modes}

\indent \indent
The 2--photon decay modes are not affected by isospin
symmetry breaking in the vector sector and keep their usual form within
the HLS model \cite{rad,mixing,chpt}~:
\begin{equation}
\left \{
\begin{array}{lll}
G_{\eta \gamma \gamma} = && 
-\displaystyle \frac{\alpha_{em}}{\pi \sqrt{3} f_{\pi}}
\left [ \frac{5-2Z}{3}\cos{\theta_P}-\sqrt{2} 
\frac{5+Z}{3}x \sin{\theta_P} \right ]\\[0.3cm] 
G_{\eta' \gamma \gamma} = && 
-\displaystyle \frac{\alpha_{em}}{\pi \sqrt{3} f_{\pi}}
\left [ \frac{5-2Z}{3}\sin{\theta_P} 
+ \sqrt{2} \frac{5+Z}{3}x \cos{\theta_P} \right ]\\[0.3cm] 
G_{\pi^0 \gamma \gamma} = && -\displaystyle  \frac{\alpha_{em}}{\pi  f_{\pi}}~~.
\end{array}
\right .
\label{brk5}
\end{equation}
As stated in the text, we actually replace this last coupling by the world average value 
for $f_{\pi}$ as given in the RPP \cite{PDG}.

Finally, the  leptonic decay widths of vector mesons depend
on the HLS $V-\gamma$ couplings. For the  ideal combinations,
we have~:
\begin{equation}
\left \{
\begin{array}{lll}
f_{\rho_I \gamma} = & \displaystyle a f_{\pi}^2 g \\[0.3cm] 
f_{\omega_I \gamma} = & \displaystyle \frac{f_{\rho_I \gamma}}{3}
\\[0.3cm] 
f_{\phi_I\gamma} = & \displaystyle  \frac{f_{\rho_I \gamma}}{3}
\sqrt{2} \ell_V ~~.
\end{array}
\right.
\label{brk6}
\end{equation}

\subsection{Partial widths}

\indent \indent
We list for completeness in this Section the expressions for the partial widths
in terms of the coupling constants for the various cases.

The two--photon partial widths are~:
\be
\Gamma(P \ra \gamma \gamma)= \displaystyle 
\frac{m_P^3}{64 \pi} |G_{P\gamma \gamma}|^2~~~,~~P=~\pi^0,~\eta,~\eta'~~.
\ee

The leptonic partial widths are~: 
\be
\Gamma(V \ra e^+ e^-)=\displaystyle \frac{4\pi\alpha^2}{3 m_V^3} |f_{V\gamma}|^2~~.
\ee

The radiative widths are~:
\be
\Gamma(V \ra P \gamma)= \displaystyle \frac{1}{96 \pi}
 \left[ \frac{m_V^2-m_P^2}{m_V} \right]^3 |G_{VP\gamma}|^2~~,
\ee
where $V$ is either of $\rho^0$, $\omg$, $\phi$ and
$P$ is either of $\pi^0$, $\eta$, $\eta'$, and~:
\be
\Gamma(P \ra V \gamma)= \displaystyle \frac{1}{32 \pi}
 \left[ \frac{m_P^2-m_V^2}{m_P} \right]^3 |G_{VP\gamma}|^2~~.
\ee

The decay width for a vector meson decaying to $V+P$ is~:
 
\be
\Gamma(V' \ra V P)= \displaystyle \frac{1}{96 \pi}
 \left[ 
 \frac{\sqrt{[m_{V'}^2-(m_V+m_P)^2][m_{V'}^2-(m_V-m_P)^2]}}{m_{V'}}
 \right]^3 |G_{V'VP}|^2~~.
\ee

Finally, the partial width for a vector meson decaying into two
pseudoscalar mesons of equal masses is~:
\be
\Gamma(V \ra P P)=\displaystyle \frac{1}{48 \pi}
 \frac{[m_V^2-4m_P^2]^{3/2}}{m_V^2} |G_{VPP}|^2~~.
\ee

\end{document}